\documentclass[12pt]{iopart}

\usepackage{iopams}
\usepackage{graphicx}
\usepackage{setspace}
\usepackage[dvips,usenames]{color}

\newcommand{\eg}{{\it e.g.}}
\newcommand{\ie}{{\it i.e.}}
\begin{document}

\title{First passages for a search by a swarm of independent random
searchers}

\author{Carlos Mej\'{\i}a-Monasterio$^1$, Gleb Oshanin$^2$ and  Gr\'egory Schehr$^3$}

\address{$^1$ Laboratory of Physical Properties,
Technical University of Madrid, Av. Complutense s/n, 28040 Madrid, Spain}
\address{$^2$Laboratoire de Physique Th\'eorique de la Mati\`ere Condens\'ee (UMR CNRS 7600), Universit\'e Pierre et Marie Curie/CNRS, 4
place Jussieu, 75252 Paris Cedex 5 France}
\address{$^3$ Laboratoire de Physique Th\'eorique (UMR CNRS 8627), Universit\'e de Paris-Sud/CNRS, France}
\ead{carlos.mejia@upm.es, oshanin@lptmc.jussieu.fr, gregory.schehr@th.u-psud.fr}

\begin{abstract}
  In this paper we study some aspects of search for an immobile target
  by a swarm of $N$ non-communicating, randomly moving searchers
   (numbered by the index $k$, $k = 1, 2, \ldots, N$), which all
  start their random motion
  simultaneously at the same point in space.
  For each realization of the search process, we record the unordered set of
  time moments $\{\tau_k\}$,
  where $\tau_k$ is the time of the first passage of the $k$-th searcher
  to the location of the target.
 Clearly, $\tau_k$'s are independent, identically
  distributed random variables with the same distribution function
  $\Psi(\tau)$.  We evaluate then the distribution $P(\omega)$ of the
  random variable $\omega \sim \tau_1/\overline{\tau}$, where
  $\overline{\tau} = N^{-1} \sum_{k=1}^N \tau_k$ is the
  ensemble-averaged realization-dependent first passage time.  We show
  that $P(\omega)$ exhibits quite a non-trivial and sometimes a
  counterintuitive behaviour. We demonstrate that in some well-studied
  cases (\eg, Brownian motion in finite $d$-dimensional domains) the
  \textit{mean} first passage time is not a robust measure of the
  search efficiency, despite the fact that $\Psi(\tau)$ has moments of
  arbitrary order. This implies, in particular, that even in this
  simplest case (not saying about complex systems and/or  anomalous
  diffusion) first passage data extracted from a single particle
  tracking should be regarded with an appropriate caution because of
  the significant sample-to-sample fluctuations.
\end{abstract}

\vspace{2pc}

\vspace{2pc}
\journal{ JSTAT}
\maketitle

\section*{Introduction}

Search processes are ubiquitous in Nature: In order to survive,
predators have to hunt the prey and the prey have to forage
\cite{stephens,bell,klafter}. In order to convert into required reaction products,
the reactants involved in chemical or biochemical reactions have first
to find each other~\cite{loverdo}.  In many biophysical processes
ligands search for binding sites \cite{net,net1}, proteins seek
the target sequences on DNA's \cite{berg}, etc.  Human beings look for
a better job, partners, shelter, files in databases. Even an attempt
to unlock a pin-protected device can be considered as a search in the
space of all possible passwords \cite{pass}.

Search for a desired target depends generally on a variety of
different conditions and may take place in different environments:
targets may be sparse, hidden, difficult to detect even when found.
The targets may be immobile or mobile, try to avoid searchers or to
evade from the searched area \cite{pnas,kam}.  They may have no or may
have their own life-time and vanish before they are detected.

Searchers may be immobile, as it happens, \eg, in visual search
\cite{visual}, may move freely or interact with the environment.
Their motion may be hindered under conditions of molecular crowding \cite{crowding},
\eg, in cell's cytoplasm or in dynamical backgrounds formed by other randomly moving particles,
or facilitated due to interactions with the
molecular motors.  The searchers may search "blindly" detecting the
target only upon an encounter with it, or "smell" (or "see") the
target somehow at long distances correcting their motion
\cite{pnas,smell}. They may have no memory of previously visited area
or adapt their strategy "on-line" repelling themselves from their
footprints on the searched substrate. Finally, the searchers may act
individually or in swarms  \cite{massimo,gelenbe,satya}.

In general, for each specific situation different search strategies
may be realized, and the question of efficient ones has motivated a
great deal of work within the last years.  While earlier works have
considered deterministic search algorithms (see, e.g.,
Refs.\cite{stephens,bell,stone} and references therein) specific to
such human activities as, say, search for natural resources or rescue
operations, more recent studies focused on random search strategies.
It was realized that the strategies based on L\'evy flights or walks
\cite{klafter,viswanathan,boyer}, in which a searcher performs
excursions whose lengths are random variables with heavy-tailed
distributions, in some aspects are more advantageous than a search based on
a conventional Brownian motion, or on random walks which step on
nearest-neighbors only. Naturally, in this case the large-scale
dynamics of searchers is superdiffusive.

Following the observation of trajectories of foraging animals in which
active local search phases randomly alternate with relocation phases
(see, e.g., Refs.~\cite{kramer}) another type of random search - an
intermittent search - has been proposed. In this algorithm the search
process is characterized by two distinct types of motion - ballistic
relocation stage when the searcher is non-receptive to the target and
a relatively slow phase with a random Brownian-type motion when the
target may be detected \cite{inter1,inter2,inter3,inter4}. Much effort
has been invested recently in understanding different optimization
schemes for such a random search. In particular, one looked for the
conditions allowing to minimize the mean first passage time for the
process which is unlimited in time \cite{inter1}, or seeked to enhance
the chances of successful detection by minimizing the non-detection
probability for the search process constrained to happen within a
finite time interval \cite{inter2,inter3,inter4}. Note that for such a
search the large-scale dynamics is diffusive, albeit intermittent.

Finally, a combination of a L\'evy-based and intermittent search has
been proposed in Ref.~\cite{combined}, in which the length of the
relocation stage was taken as a random variable with a heavy-tailed
distribution. It was shown that such a combined strategy is
advantageous in the critical case of rare targets.

\begin{figure}[ht]
  \centerline{\includegraphics*[width=0.45\textwidth]{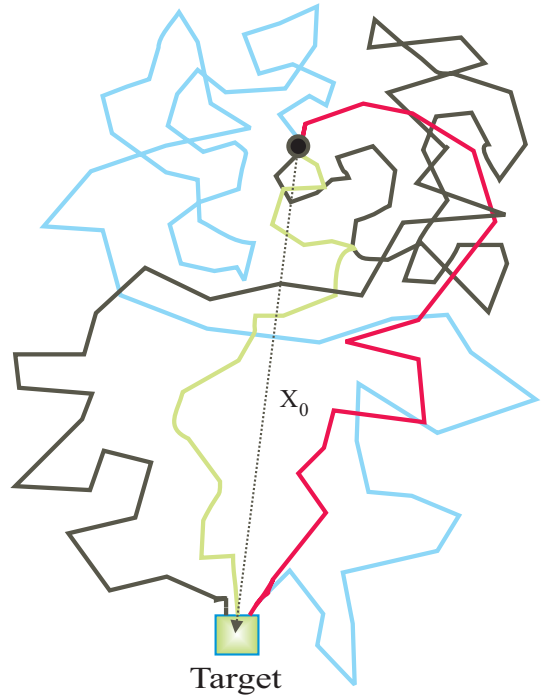}}
  \caption{A sketch of trajectories of four searchers starting at the
    same point (filled black circle) at a distance $x_0$ from the target
    and reaching the target for the first time at different
    time moments.}
  \label{target}
\end{figure}

In this paper we discuss some aspects of a blind search by a swarm of
$N$ independent, \textit{non-communicating} searchers. We consider a
situation, as depicted in Fig.~\ref{target}, in which $N$ searchers  (numbered by the index $k$, $k = 1, 2, \ldots, N$)
occupy initially the same position in space, at some distance $x_0$
apart of an immobile target, start their random motion simultaneously
and arrive for the first time to the location of the target at times
$\tau_k$, respectively. Note that $\tau_k$s are not ordered. Clearly, in such a
situation the first passage times $\tau_k$'s are independent,
identically distributed random variables with the same distribution
$\Psi(\tau)$.

We focus here on the random variable
\begin{equation}
\omega = \frac{1}{N} \, \frac{\tau_1}{\overline{\tau}}, \label{def}
\end{equation}
where $\overline{\tau}$ is the averaged, over the ensemble of $N$
searchers, realization-dependent first passage time,
\begin{equation}
\overline{\tau} = \frac{1}{N} \, \sum_{k = 1}^N \tau_k.
\end{equation}
Hence, the random variable $\omega$ probes the first passage time of a
given searcher relative to the ensemble-averaged first passage time
for $N$ independent searchers.  The scaling factor $1/N$ in
Eq.~(\ref{def}) is introduced here for convenience, so that regardless
of the value of $N$, the random variable $\omega$ has a support on
$[0,1]$. We note parenthetically that random variables such as in
Eq.~(\ref{def}) were previously studied in
Refs.~\cite{iddo1,sid,iddo2,iddo3,greg1,greg2} within a different
context.

Our goal is to calculate the distribution function
\begin{equation}
P(\omega) = \Big \langle \delta\left(\omega - \frac{1}{N} \, \frac{\tau_1}{\overline{\tau}}\right)\Big \rangle \;,
\end{equation}
where $\delta(\cdot)$ is the delta-function and the angle brackets denote
averaging over different realizations of independent, identically
distributed random variables $\tau_k$.

For arbitrary normalized $\Psi(\tau)$, the distribution $P(\omega)$ is
normalized, possesses all moments and the first moment $\langle \omega
\rangle \equiv \int^1_0 \omega \, d\omega \, P(\omega) = 1/N$. Note that
$P(\omega)$ can be seen as a measure of the robustness of a given search
algorithm. Clearly, if $P(\omega)$ appears to be sharply peaked at
$\omega = 1/N$, this would signify that the underlying search
algorithm is quite robust.  Otherwise, if the distribution appears to
be broad, or even to have a multi-modal shape, this would imply that
the performance of such an algorithm is rather poor and
sample-to-sample fluctuations matter.  From yet another conceptual
perspective, one can say that $P(\omega)$ probes the validity of the
\textit{mean} first passage time as the proper measure of a search
process efficiency.  We proceed to show that in many situations $P(\omega)$ has a
rather complicated structure of which the mean behavior is not
representative.  We will show that in different situations $P(\omega)$
may have completely different shapes (modality) and also change the
shape (say, from a unimodal bell-shaped form to a bimodal $U$- or an
$M$-shaped one) when some of the parameters are slightly modified. A
similar phenomenon of a shape reversal has been previously observed
for different mathematical objects in Refs.~\cite{sid,greg1,greg2}.

We turn next to the key feature: the distribution $\Psi(\tau)$ of
$\tau_k$'s, which encodes all the information on the specific
properties of the searchers' random motion and on their initial location
relative to the target. Without a significant lack of generality, we
suppose that for a search in infinitely large systems, in which all
searchers perform a blind search and detect the target with
probability $1$ upon a first encounter with it, $\Psi(\tau)$ can be
defined as
\begin{equation}
\label{distribution}
 \Psi(\tau) = \frac{a^{\mu}}{\Gamma(\mu)} \, \exp\left(- \frac{a}{\tau}\right) \, \frac{1}{\tau^{1 + \mu}} ,
\end{equation}
where $\Gamma(\cdot)$ is the gamma function, $a$ is a characteristic
parameter which sets the cutoff of the distribution at small values of
$\tau$, and $\mu \geq 0$ is the so-called the persistence exponent
\cite{satya_review_persistence}. Note that depending on the
dimensionality of space and type of random motion, one can encounter
completely different values of $\mu$; $\mu$ can be $0 \leq \mu < 1$,
$\mu = 1$ or $\mu > 1$. The distribution in Eq.~(\ref{distribution}) is normalized
but does not possess already a first moment, (i.e., the mean first passage time), for $0 \leq \mu < 1$.

The distribution in Eq.~(\ref{distribution}) is exact for a Brownian
motion (BM) in semi-infinite one-dimensional (1D) systems, in which case
$\mu = 1/2$ and $a = x_0^2/4 D$, $D$ being the diffusion coefficient
\cite{redner}.  In fact, the case $\mu = 1/2$ appears to be very
representative.  According to the theorem due to Sparre Andersen
\cite{sparre}, in 1D systems for any discrete-time random
walk with each step length chosen from a continuous, symmetric but
otherwise arbitrary distribution the first passage time distribution
$\Psi(n)$ decays with a number of steps $n$ as $n^{-3/2}$. For
continuous-time $\tau$ Markov processes, an analogous result is
$\Psi(\tau) \sim \tau^{-3/2}$. This universality, of course, is broken when subordination effects
(i.e., long
tailed waiting time distributions in a CTRW sense) are present.

Although not exact, the distribution in Eq.~(\ref{distribution}) is
physically quite plausible for other types of random motion in
infinite 1D systems provided that an appropriate choice
of the exponent $\mu$ is made \cite{redner}: In particular,
Eq.~(\ref{distribution}) with $\mu = 1 - 1/\alpha$ is a reasonable
approximation for the first-arrival probability density for L\'evy
flights with L\'evy index $\alpha$, $1 < \alpha < 2$ \cite{chechkin}
and with $\mu = 1 - H$ \cite{molchan} - for the first-passage-time
distribution for fractional BM with Hurst index $H$, $0 < H < 1$.  One
may also claim that $\mu = 1 - d_f/d_w$ for $d_f < d_w$ (compact
exploration \cite{pgg}), where $d_f$ is a non-integer spatial
dimension and $d_w$ is the fractal dimension of random motion
trajectories, since the first passage distribution is defined as the
time derivative of the survival probability of an immobile target or
of the normalized current through the surface of the
target~\cite{redner}. Hence, its long-time tail is the same as the
long-time tail of the time derivative of the Smoluchowski constant
\cite{osh0,osh1}.

The case $\mu = 0$ can be encountered in border-line situations of
compact exploration, when the dimension of space $d_f$ equals the
fractal dimension $d_w$ of the random motion trajectories. In
particular, this situation is realized for standard BM ($d_w = 2$) in
two-dimensional (2D) space. Here, one finds that the \textit{long-time}
tail of the distribution $\Psi(\tau)$ follows (see, \eg,
Ref.\cite{redner})
\begin{equation}
\label{distribution_log}
 \Psi(\tau) \sim \frac{1}{\tau \ln^2(\tau)}.
\end{equation}
This situation will be considered in more detail in what follows.

Finally, one finds $\mu = d/2 - 1$, (or, more generally, $\mu = d_f/d_w
- 1$), for Brownian (anomalous) motion in $d > 2$ ($d_f > d_w$)
dimensional systems. Note, however, that here the expression in
Eq.~(\ref{distribution}) is not normalized and thus can not be
considered as a distribution since only a finite fraction of
trajectories will visit the target within an infinite time so that the
searchers will have a finite probability to escape to infinity (non
compact exploration \cite{pgg}). One can, however, normalize
$\Psi(\tau)$ in Eq.~(\ref{distribution}) by hand, treating it as the
conditional probability distribution of the first passage times for
such trajectories which visit the target within an infinite time.

Next, we will consider a normalized, exponentially-truncated version of
the distribution in Eq.~(\ref{distribution}):
\begin{equation}
\label{truncated_dist}
 \Psi(\tau) = \frac{\left(a b\right)^{\mu/2}}{2 K_{\mu}(2 \sqrt{a/b})} \, \exp\left(- \frac{a}{\tau}\right) \, \frac{1}{\tau^{1 + \mu}} \, \exp\left(- \frac{\tau}{b}\right),
\end{equation}
where $K_{\mu}(\cdot)$ is the modified Bessel function. In contrast to
$\Psi(\tau)$ in Eq.~(\ref{distribution}), the distribution in
Eq.~(\ref{truncated_dist}) possesses moments of arbitrary order. Note
that the latter point is crucial and, according to a common belief, the
first moment of the distribution, \ie, the \textit{mean} first passage
time can be regarded as a robust measure of the search process efficiency. We set
out to show that in many situations this is not the case due to the
significant sample-to-sample fluctuations.

The distribution in Eq.~(\ref{truncated_dist}) is exact for
a BM in semi-infinite 1D systems in presence of a constant
bias pointing towards the target (see, \eg, Ref.\cite{redner}). In
this case one has $a = x_0^2/4 D$, where $x_0$ is the starting point
and  $b = 4 D/v^2$, $v$ being the
drift velocity.

One may argue, as well, that the expression in
Eq.~(\ref{truncated_dist}) is an appropriate approximation for the
first-passage time distribution for search in finite systems or search
assisted by smell. In general, of course, for random motion in finite
systems the distribution $\Psi(\tau)$ will be represented as a series
of exponentials; the form in Eq.~(\ref{truncated_dist}) is thus
tantamount to a heuristic approximation of this series in which one
takes a behavior specific to an infinite system and truncates it by an
exponential function with the characteristic decay time equal to the
largest relaxation time. Depending on a particular situation, the
cut-off parameter $b$ will be either proportional to the volume of the
system (for BM in $d > 2$ systems), to $L^2$ where $L$ is the length
of the interval for BM in 1D, to $S \ln(S) $ for BM in
finite 2D systems of area $S$, or inversely proportional
to the strength of the bias for biased diffusion. Some of these
situations will be discussed in detail in the sequel.

We finally remark that the distribution in Eq.~(\ref{truncated_dist})
appears in many other physical problems. To name but a few we mention
the distribution of times between action potentials (or the ISI
distribution) in the integrate-and-fire model of neuron dynamics
\cite{man}, the distribution of the stopping distances for the sliding
motion of a solid block on an inclined heterogeneous plane
\cite{lima}, the avalanche life-time distribution in the mean-field
version of the Bak-Sneppen model \cite{flyv}, the distribution of the
probability current in finite disordered one-dimensional samples
\cite{osh} or the distribution of the number of times that a particle
diffusing in a sphere hits its boundary during some time interval
\cite{net1}.  Thus our subsequent analysis applies to these systems as
well.

The outline of our paper is as follows: We start in section 1 with a
general remark on the tails of the first passage time distribution in
case with $N$ independent randomly moving searchers. In section 2 we
derive an exact result for the distribution $P(\omega)$ for $\Psi(\tau)$ in Eq.~(\ref{distribution}) with arbitrary
$\mu$ and for arbitrary $N$. In section 3 we discuss several exactly solvable
particular cases.  Section 4 is devoted to the asymptotic analysis of
$P(\omega)$ for large $N$ and arbitrary $\mu$. In section 5 we
evaluate $P(\omega)$ for the exponentially-truncated distribution in
Eq.~(\ref{truncated_dist}) in case of two searchers. We also furnish
here exact calculations of $P(\omega)$ for two BMs in finite 1D, 2D and 3D
spherical domains. Next, in section 6, we discuss the form
of $P(\omega)$ for $N$ searchers whose first passage time distribution
is given by Eq.~(\ref{truncated_dist}).  Finally, in section 7 we
conclude with a summary of our results and some generalizations.

\section{First passage to the target for $N$ independent searchers}

We begin with a somewhat evident but conceptually very important
remark on the tails of the first passage time distribution in case of
$N$ independent, non-communicating randomly moving searchers.  For the
original results and a discussion we address the reader to
Ref.~\cite{katja}. Some other interesting aspects of this model were discussed in Refs.~\cite{gelenbe,paul0,paul}.

Let $P_N(\tau)$ denote the probability that up to time moment $\tau$
neither of $N$ searchers has visited the target, \ie, the target
remained non-detected up to time $\tau$. For the situation under
consideration, clearly,
\begin{equation}
P_N(\tau) = P_1(\tau)^N,
\end{equation}
where $P_1(\tau)$ is an analogous probability for a single
searcher. Supposing that the target is found as soon as any of the
searchers arrives to its location for the first time, we have that for
$N$ searchers the first passage density $\Psi_N(\tau)$ is given by
\begin{equation}
\label{qq}
\Psi_N(\tau) = - \frac{d P_N(\tau)}{d \tau}.
\end{equation}
This yields immediately that $\Psi_N(\tau)$ follows (for the parental distribution in Eq.~(\ref{distribution})), as $\tau \to
\infty$,
\begin{equation}
\label{dist1}
\Psi_N(\tau) \sim \frac{1}{\tau^{1 + \mu N}}.
\end{equation}
A remarkable feature of this simple result is that depending on the
number of searchers $N$, the distribution in Eq.~(\ref{dist1}) may
have finite moments even if $\mu \leq 1$, contrary to the parent
distribution in Eq.~(\ref{distribution}) which does not have any
moment for such values of $\mu$. For instance, for $0 < \mu < 1$ and
$\mu N > 1$, Eq.~(\ref{dist1}) has a finite first moment, \ie, a
finite mean first passage time, for $\mu N > 2$ it has a finite second
moment, and generally, for $\mu N > k$ it has $k$ first finite
moments. This is a crucially important advantage of search processes
involving $N$ searchers, which makes the search process more efficient
even for non-communicating searchers.

Next, it might be instructive to consider some exactly solvable case,
\ie, BM in semi-infinite 1D systems, and to calculate the
$N$-dependence of the mean first passage time. In this case, the
probability that a single searcher has not visited the location of the
target (the origin) up to time moment $\tau$, starting at distance
$x_0$ from the target, obeys $P_1(\tau) = {\rm erf}\left(x_0/\sqrt{4 D
    \tau}\right)$, where ${\rm erf}(.)$ is the error function.  Hence,
the first moment of the distribution in Eq.~(\ref{qq}), \ie, the mean
first passage time $\big<\tau_N\big>$ for an ensemble of $N$ independent
searchers is defined as
\begin{equation}
\big< \tau_N \big> = - \int^{\infty}_0 d\tau \, \tau \, \frac{d P_N(\tau)}{d \tau} = \int^{\infty}_0 d\tau \, {\rm erf}^{N}\left(\frac{x_0}{\sqrt{4 D \tau}}\right).
\end{equation}
The integral in the latter equation is convergent, as we have already
remarked, for $N \geq 3$. One verifies that $\big< t_N \big>$ is a
slowly \textit{decreasing} function of the number of searchers $N$,
and, for $N \gg 1$, we find that
\begin{equation}
\label{t}
\big< \tau_N \big> \sim \frac{x_0^2}{4 D \ln\left(N\right)}.
\end{equation}
Therefore, the more searchers one has, the less the mean first passage
time is.  Curiously enough, this essentially 1D result,
$\big< \tau_N \big>$ in Eq.~(\ref{t}), coincides exactly with the mean
residence time which $N$ BMs spend simultaneously together in a
circular disc of radius $x_0$ on a 2D plane within an
infinite time interval \cite{occupation}, or with the mean first exit
time of one of $N$ BMs from such a disc \cite{yuste}.

Note finally that the mean first passage time may acquire a much
stronger dependence on $N$ and decrease much faster for communicating
searchers which share the information on the location of the target
\cite{massimo}.  Note, as well, that increasing the number of
"searchers" allows to decrease substantially the time necessary to
reach the target site on a DNA for proteins which strongly bind to
other nonspecific sites acting as deep temporal traps \cite{raphael}.
Some properties of the first passage time distribution for the event
in which $N$ random walks appear for the first time simultaneously at
the same lattice site have been discussed in Ref.~\cite{pnas} within
the context of a survival of an evasive prey.

\section{General form of $P(\omega)$ for heavy-tailed first passage time distributions}

Let $\Big< \exp\left( - \lambda \omega\right)\Big>$, $\lambda \geq 0$,
denote the moment generating function of the random variable $\omega$,
Eq.~(\ref{def}).  For arbitrary $\Psi(\tau)$, it can be formally
represented as an $N$-fold integral:
\begin{equation}
 \left< e^{- \lambda \omega}\right> = \int^{\infty}_0 \ldots \int^{\infty}_0 \left( \prod_{n = 1}^N d \tau_n \, \Psi(\tau_n)\right)  \,
\exp\left(-\lambda \frac{\tau_1}{\tau_1 + \tau_2 + \ldots + \tau_N}  \right).
\end{equation}
Integrating over $d \tau_1$, we change the integration
variable $\tau _1 \to \omega$, to get
\begin{eqnarray}
\left< e^{- \lambda \omega}\right> &=& \int^1_0 \frac{d \omega}{\left(1 - \omega\right)^2} \, e^{- \lambda \omega}
\int^{\infty}_0 \ldots \int^{\infty}_0 \left( \prod_{n = 2}^N d \tau_n \, \Psi(\tau_n)\right)  \nonumber\\
&\times& \left(\tau_2 + \ldots + \tau_N \right) \Psi\left(\frac{\omega}{1 - \omega} \left(\tau_2 + \ldots + \tau_N \right) \right).
\end{eqnarray}
Using next the following integral representation
\begin{equation}
 \tau \Psi(\tau) = \int^{\infty}_0 dp \, Q(p) \; e^{ - p \tau},
\end{equation}
where the kernel $Q(p)$ is some unknown function defined via the
inverse Laplace transform of the distribution $\Psi(\tau)$, we obtain
the following general result for the probability density $P(\omega)$
in case of $N$ identic $\tau$-variables:
\begin{eqnarray}
\label{general}
 P(\omega) = \frac{1}{\omega^2} \int^{\infty}_0 d\lambda \, Q\left(\frac{1 - \omega}{\omega} \lambda\right) \; \Phi^{N - 1}\left(\lambda\right),
\end{eqnarray}
with $\Phi(\lambda) = \left< \exp(- \lambda \tau)\right>$ being the characteristic  function of the distribution $\Psi(\tau)$.

Now,  for the distribution in Eq.(\ref{distribution}) we have
\begin{equation}
\label{1}
Q(p) = \frac{a^{(\mu + 1)/2}}{\Gamma(\mu)} p^{(\mu - 1)/2} J_{\mu - 1}\left(2 \sqrt{a p}\right),
\end{equation}
and
\begin{equation}
\label{2}
\Phi(\lambda) = \frac{2 a^{\mu/2}}{\Gamma(\mu)} \lambda^{\mu/2} K_{\mu}\left(2 \sqrt{a \lambda}\right),
\end{equation}
where $J_{\nu}(\cdot)$ is the Bessel function.

Substituting the expressions in Eqs.~(\ref{1}) and (\ref{2}) into
Eq.~(\ref{general}), we find that the probability density $P(\omega)$
is given by
\begin{eqnarray}
\label{m} P(\omega) = \frac{2^{-
\mu}}{\Gamma(\mu)}  \frac{\left(1 - \omega\right)^{(\mu - 1)/2}}{\omega^{(3 + \mu)/2}} \,
\int^{\infty}_0 du \, u^{\mu } \, J_{\mu - 1}\left(\sqrt{\frac{1 - \omega}{\omega}} \, u\right) \, \Xi^{N-1}(u),
\end{eqnarray}
independently of $a$, with
\begin{equation}
\label{K} \Xi(u) = \frac{2^{1 - \mu} }{\Gamma(\mu)} \, u^{\mu} \,
{K}_{\mu}\left( u\right).
\end{equation}
The result in Eq.~(\ref{m}) defines an exact distribution $P(\omega)$
for arbitrary $\mu$ and $N$.  In several particular cases, the
integral in Eq.~(\ref{m}) can be performed in closed form: when $N =
2$ or $N = 3$ and arbitrary $\mu > 0$, or when $\mu$ is equal to a
half of an odd integer, while $N$ is arbitrary. We discuss below some
of these cases, as well as present an asymptotic analysis of
$P(\omega)$ in the limit $N \gg 1$.

\section{Exactly solvable cases for the heavy-tailed distributions}

\subsection{Two non-communicating random searchers}

Consider first the case of just two random searchers, $N = 2$.  For
this simple situation, one readily finds from Eq.~(\ref{m}) that
$P(\omega)$ is given explicitly by:
\begin{equation}
\label{lim}
P(\omega) = \frac{\Gamma(2 \mu)}{\Gamma^2(\mu)} \, \omega^{\mu - 1} \left(1 - \omega\right)^{\mu - 1},
\end{equation}
\ie, in this case $P(\omega)$ is a beta-distribution.  This result has
been also obtained within a different context in Ref.~\cite{iddo1}.

\begin{figure}[ht]
  \centerline{\includegraphics*[width=0.65\textwidth]{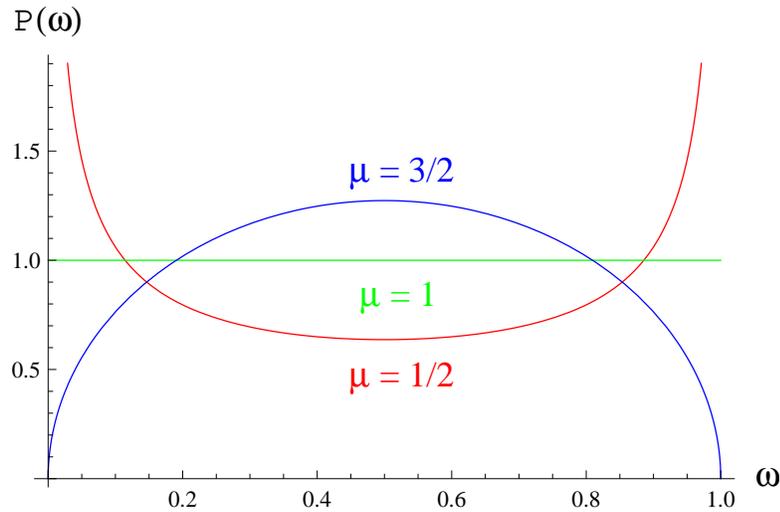}}
\caption{The distribution $P(\omega)$ in Eq.~(\ref{lim}) for $\mu = 1/2$, $\mu = 1$ and $\mu = 3/2$.}
  \label{fig2}
\end{figure}

Notice now that, despite its simplicity, the result in Eq.~(\ref{lim})
contains a surprise: it has a completely different shape (modality)
depending on whether $0 < \mu < 1$, $\mu = 1$ or $\mu > 1$ (see
Fig.~\ref{fig2}).
When $0 < \mu < 1$, the distribution $P(\omega)$ has a characteristic
$U$-shape, so that the most probable values of $\omega$ are $0$ and
$1$.  Strikingly, the mean $\big< \omega\big> = 1/2$ corresponds in
this case to the \textit{least probable} value of the distribution.
This signifies that in this case there is no symmetry between two
identical searchers and both arrive to the target for the first time
at distinctly different times.  Note that for $\mu = 1/2$ (two BMs in
1D), $P(\omega) = 1/\pi \sqrt{\omega (1 - \omega)}$ and hence, the
probability $P(\omega \leq \Omega)$ that $\omega$ attains any value
from the interval $[0, \Omega]$ obeys
\begin{equation}
P(\omega \leq \Omega) = \frac{2}{\pi} {\rm arcsin}\left(\sqrt{\Omega}\right),
\end{equation}
which is the continuous arcsine distribution.

When $\mu = 1$, the distribution $P(\omega)$ in Eq.~(\ref{lim}) is
uniform, which means that for either of the searchers its first
passage time, relative to the average over the ensemble of two
searchers, may take any value with equal probability.

Finally, for $\mu > 1$ the distribution $P(\omega)$ is unimodal and
centered at $\omega = 1/2$. This signifies that in this case two
searchers will most likely arrive to the target simultaneously.

Therefore, in situations when $0 < \mu \leq 1$ (\ie, when $\Psi(\tau)$
does not have already the first moment) the first passage times of two
searchers will be most probably distinctly different. On contrary, for
$\mu > 1$ the first passage times will be most probably the same and
hence, the algorithm resulting in such values of $\mu$ will be robust.

We end up this subsection by noticing that a similar transition was
found in Ref.~\cite{sanjib_sinai} in the related Sinai model with a
linear drift of strength $\mu$. There the occupation time distribution
on the positive axis exhibits a transition at $\mu=1$ and is also
given, in certain limiting cases, by a beta-distribution, as in
Eq.~(\ref{lim}).

\subsection{Three non-communicating random searchers}

\begin{figure}[ht]
  \centerline{\includegraphics*[width=0.65\textwidth]{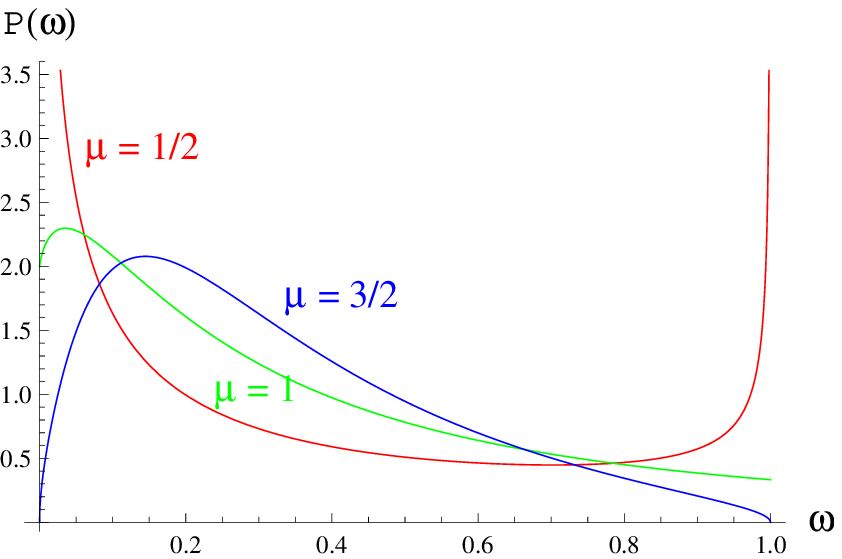}}
\caption{The distribution $P(\omega)$ in Eq.~(\ref{N3}) for $\mu = 1/2$, $\mu = 1$ and $\mu = 3/2$.}
  \label{fig3}
\end{figure}

For three non-communicating searchers, whose first passage times obey
a non-truncated distribution in Eq.~(\ref{distribution}), we find
\begin{eqnarray}
\label{N3}
P(\omega) &=& \frac{\sqrt{\pi}}{2^{4 \mu - 1}} \; \frac{\Gamma(2 \mu) \Gamma(3 \mu)}{\Gamma^3(\mu) \Gamma(2 \mu + 1/2)}  \omega^{-1 - \mu} \left(1 - \omega\right)^{\mu - 1}  \nonumber\\
&\times& \;_2F_1\left(2 \mu,
3 \mu; 2 \mu +\frac{1}{2}; - \frac{1 -  \omega}{4 \omega}\right),
\end{eqnarray}
where $_2F_1$ is a hypergeometric series. The distribution in
Eq.~(\ref{N3}) for three different values of $\mu$ is depicted in
Fig.~\ref{fig3}.

Consider first behavior of the distribution in Eq.~(\ref{N3}) in the
vicinity of $\omega = 0$ and $\omega = 1$.  Recollecting the
definition of the hypergeometric series, one immediately observes from
Eq.~(\ref{N3}) that $P(\omega) \sim C_2 (1-\omega)^{\mu - 1}$ when
$\omega \to 1$. A little bit more involved analysis shows that
$P(\omega) \sim C_1 \omega^{\mu - 1}$ when $\omega \to 0$. This means
that similarly to the $N = 2$ case, the distribution $P(\omega)$
diverges at both edges when $\mu < 1$, and most probable values of
$\omega$ are $0$ and $1$.  Note, however, that $C_1 > C_2$ and hence,
the distribution is skewed to the left favoring small values of
$\omega$.  Therefore, for $\mu < 1$ the most probable situation is
that one of three searchers arrives to the location of the target much
earlier than two others. Clearly, the mean $\langle \omega \rangle =
1/3$ does not have any significance [apart, of course, of the fact that
this is just the first moment of the distribution in Eq.~(\ref{N3})].

Further on, the distribution in Eq.~(\ref{N3}) exhibits a qualitative
change of behavior for $\mu \geq 1$. Here $P(\omega)$ is always a
bell-shaped function of the variable $\omega$ centered at the most probable value
$\omega = \omega_m$. The only difference between the $\mu = 1$ and
$\mu > 1$ cases is that for the former $P(\omega)$ attains a non-zero
values at the edges, $P(\omega = 1) = 1/3$ and $P(\omega=0) = 2$,
while in the latter $P(\omega = 0) = P(\omega = 1) \equiv 0$.  It is
important to observe, however, that $\omega_m$ is always appreciably
\textit{less} than $\big< \omega \big> = 1/3$. For example, for $\mu
=3$, one has $\omega_m \approx 0.2719$, for $\mu = 10$ one has
$\omega_m \approx 0.3102$ and etc. In fact, $\omega_m \to 1/3$ only
when $\mu \to \infty$.

\subsection{$N$ non-communicating random searchers for $\mu = 1/2$ and $\mu = 3/2$}

\begin{figure}[ht]
  \centerline{\includegraphics*[width=0.65\textwidth]{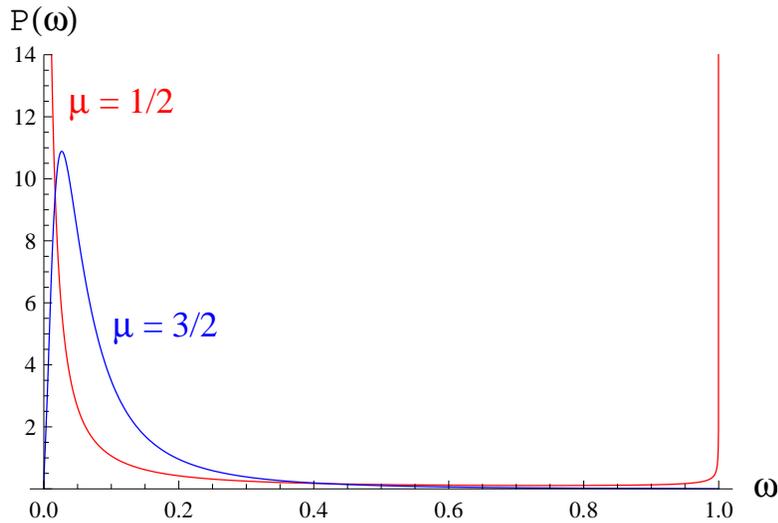}}
\caption{The distribution $P(\omega)$ in Eqs.~(\ref{mu12}) (red line) and (\ref{mu32}) (blue line) for $N = 10$.}
  \label{fig4}
\end{figure}

For $\mu = 1/2$ and arbitrary $N$, (\ie, for $N$ BMs starting from the
same point on a semi-infinite line), we find the following simple law:
\begin{equation}
\label{mu12}
 P(\omega) = \frac{N - 1}{\pi} \frac{1}{\sqrt{\omega \left(1 - \omega\right)}} \frac{1}{1 - \omega + (N - 1)^2 \omega},
\end{equation}
while for
 $\mu = 3/2$ and arbitrary $N$ we get
\begin{eqnarray}
\label{mu32}
\fl P(\omega) = \frac{\Gamma(N)}{\pi} \frac{\left(1  - \omega\right)^{1/2}}{\omega \left(1 - \omega + (N - 1)^2 \omega\right)^{3/2}} \sum_{p=0}^{N-1} \frac{p + 1}{\Gamma(N - p)}  \nonumber\\
\times \left(\frac{\omega}{1 - \omega + (N-1)^2 \omega}\right)^{p/2} {\rm U}_{p + 1}\left( \left(\frac{\omega (N - 1)^2}{1 - \omega + (N-1)^2 \omega}\right)^{1/2}\right),
\end{eqnarray}
where ${\rm U}_k(\cdot)$ are the Chebyshev's polynomials of the second
kind. These distributions, for a particular case $N = 10$, are
depicted in Fig.~(\ref{fig4}).

Note that for $\mu = 1/2$ and arbitrary $N$, the distribution
$P(\omega)$ has a characteristic $U$-shape and diverges at the edges
$\omega = 0$ and $\omega = 1$. The distribution function is strongly
skewed towards the left edge; for $\omega \to 0$ the distribution
behaves as $P(\omega) \sim (N - 1) \, \omega^{-1/2}$, while for
$\omega \to 1$ we have $P(\omega) \sim (N - 1)^{-1} \, (1 -\omega)^{-
  1/2}$, i.e., the amplitudes are $(N - 1)^2$ times different.  This
means that for $N \gg 1$ an event in which one out of a swarm of $N$
independent searchers arrives to the target much earlier than the rest
of searchers is much more probable than an event in which one of the
searchers arrives to the target location much later than others.

For $\mu = 3/2$ the distribution $P(\omega)$ is unimodal and peaked at
some value $\omega = \omega_m$, which defines the most probable first
passage time of a given searcher relative to the ensemble average
value.  Note, however, that $\omega_m$ is always less than the mean
$\big< \omega \big> = 1/N$. Actually, as one may observe from
Fig.~\ref{fig4}, $\omega_m$ is more than two times less than
$1/10$. Note that in this case the mean first passage time for an
individual searcher exists, but apparently is not a proper measure of
the search process since it is different from the most probable value.

\section{Asymptotic large-$N$ behavior of $P(\omega)$ for heavy-tailed first passage time distributions}

We consider next the large-$N$ asymptotic behavior of the probability
density $P(\omega)$ in Eq.~(\ref{m}).  To do this, it might be helpful
first to look at the Eq.~(\ref{m}) from a different perspective and to
realize that $P(\omega)$ can be expressed as the probability density
function $P_{N-1}(|{\bf r}|)$ for the position of a random walker in
some Rayleigh's-type random flight model with a variable flight length
in which $N$ - (the number of independent searchers) - will play the
role of "time".

Consider a Rayleigh's random flight process (see, e.g.,
Ref.~\cite{hughes}) in a $d = 2 \mu$-dimensional continuum. If $d$ is
thought to be an integer, this would restrict the analysis to some
particular values of $\mu$. However, as noted in Ref.~\cite{kingman},
formal considerations can be used to define an isotropic random flight
in a space of arbitrary positive dimension $d$, with $d$ not
necessarily restricted to integral values. This relaxes any constraint
on the value of $\mu$.

Suppose that flights lengths are independent, identically distributed
random variables and let the probability density function $p({\bf
  r},{\bf r'})$ for a transition from ${\bf r}$ to ${\bf r'}$ be a
function of $\rho = |{\bf r} - {\bf r'}|$ only, which means that the
process is translationally invariant and $p({\bf r},{\bf r'}) =
p(\rho)$. Choose next
\begin{equation}
\label{step} p(\rho) = \frac{\Gamma(2 \mu)}{\pi^{\mu}
\Gamma(\mu)} \frac{1}{\left(1 + \rho^2\right)^{2 \mu}},
\end{equation}
\ie, a heavy-tailed (but normalizable for any $\mu > 0$)
distribution.

Using Fourier transform technique (see, \eg, Ref.~\cite{hughes}), one
readily finds that the probability density function of a random
walker, starting at the origin, to be at a distance $\rho$ from the
origin after $N - 1$ such flights, is given by
\begin{equation}\label{rw}
\label{ray} P_{N-1}(\rho) = \frac{1}{2^{\mu} \pi^{\mu} \rho^{\mu - 1}} \; \int^{\infty}_0 du \, u^{\mu } \, J_{\mu -
1}\left(\rho \, u\right) \, \Xi^{N-1}(u),
\end{equation}
with $\Xi(u)$ defined by Eq.~(\ref{K}).

Consequently, we find the following relation between the probability
density $P(\omega)$ in Eq.~(\ref{m}) and $P_{N-1}(\rho)$ in
Eq.~(\ref{ray}):
\begin{equation}\label{rel_rw}
\omega \left(1 - \omega\right) P(\omega) =
\left. \frac{\pi^{\mu}}{\Gamma(\mu)} \rho^{2 \mu} P_{N-1}(\rho)\right|_{\rho = \sqrt{(1 - \omega)/\omega}} \, .
\end{equation}
Therefore, the probability density of the random variable $\omega$,
which describes the realization-dependent ratio of the first passage
time of a given searcher in an ensemble of $N$ ones, and of the
ensemble averaged first passage time, is proportional to the
probability density of finding a random walker performing Rayleigh's
random flights with a broad distribution of flight length,
Eq.~(\ref{step}), at distance $\sqrt{(1 - \omega)/\omega}$ away from the
origin after $N - 1$ flights.

When $N$ is large, the integral in Eq.~(\ref{m}) is dominated by the
behavior of $\Xi(u)$ in the vicinity of $u = 0$. In turn, the latter
depends on the value $\mu$: the cases $\mu > 1$, $\mu =1$ and $\mu <
1$ need to be considered separately.

\subsection{The case $\mu > 1$}

In this case, the leading small-$u$ behavior of the characteristic
function follows:
\begin{equation}
\label{u}
\ln\left(\Xi(u)\right) \sim -
u^2/4 (\mu - 1) \;.
\end{equation}
Plugging the latter asymptotic form into Eq.~(\ref{m}), and performing
integration over $du$, we find that the asymptotic large-$N$ behavior
of the distribution $P(\omega)$ is determined by
\begin{equation}
\label{form}
P(\omega) \sim \frac{1}{\omega (1 - \omega)} \left(\frac{1- \omega}{\omega N}\right)^{\mu} \exp\left(- (\mu - 1) \frac{(1 - \omega)}{\omega N}\right).
\end{equation}
Therefore, in the limit $N \gg 1$ for $\mu > 1$ the distribution
$P(\omega)$ is always a bell-shaped function of $\omega$, which
approaches $0$ exponentially fast when $\omega \to 0$ and as a
power-law when $\omega \to 1$. The maximum of $P(\omega)$ is located
at
\begin{eqnarray}
\omega_m &=& \frac{1}{4 N} \left(\mu - 1 + (\mu + 1) N - \sqrt{\left(\mu - 1 + (\mu + 1) N\right)^2  - 8 (\mu - 1) N } \right) \nonumber\\ &\sim& \frac{\mu - 1}{\mu + 1} \frac{1}{N}.
\end{eqnarray}
This substantiates our previous claims that $\omega_m$ is always less
than $\big< \omega \big> = 1/N$ and converges to $1/N$ only when $\mu
\to \infty$. Note that in this case the first moment of $\Psi(\tau)$,
\ie, the mean first passage time, exists but it is not representative
of the most probable behaviour.

It is instructive now to reproduce the result in Eq.~(\ref{form})
using a different type of argument. Notice that $P_{N}(\rho)$ in
Eq.~(\ref{rw}) in the limit $N \to \infty$ becomes a Gaussian
distribution in $d = 2 \mu$ dimensional space of one rescaled variable
$\rho/N^{1/2}$:
\begin{eqnarray}
P_{N}(\rho) \sim \frac{1}{N^{\mu}} \exp{\left(-(\mu-1) \frac{\rho^2}{N}\right)}.
\end{eqnarray}
Inserting the latter expression into the relation in Eq. (\ref{rel_rw}), we recover
the result in Eq.~(\ref{form}).

\subsection{The case $\mu = 1$}

In this borderline case one finds
\begin{equation}\label{Ximu1}
\ln\left(\Xi(u)\right) \sim - \left(1 - 2 \gamma + 2 \ln 2 - 2 \ln(u) \right) u^2/4 \;,
\end{equation}
where $\gamma$ is the Euler constant.

Note that here one has an additional logarithmic factor $\ln(u)$, as
compared to the leading small-$u$ behavior in Eq.~(\ref{u}). Since
logarithm is a slowly varying function, we can repeat essentially the
same argument: $P_{N}(\rho)$ in Eq.~(\ref{rw}) converges, as $N \to
\infty$, to a Gaussian distribution of the scaling variable $\rho/({N
  \log N})^{1/2}$:
\begin{equation}
P_{N}(\rho) \sim \frac{1}{N \ln(N)} \exp\left(- \frac{\rho^2}{N \ln(N)}\right).
\end{equation}
Hence, in virtue of the relation in Eq. (\ref{rel_rw}),  we find
\begin{eqnarray}
P(\omega) \sim \frac{1}{\left(N \ln(N) \omega\right)^2} \exp\left(- \frac{(1 - \omega)}{N \ln(N) \omega}\right)\;.
\end{eqnarray}
This is again a bell-shaped function of $\omega$ with the most
probable value $\omega_m~\sim~1/(2 N \log N)$.

\subsection{The case $\mu < 1$}

For $\mu <1$ the leading small-$u$ behavior of the characteristic
function reads:
\begin{equation}
\ln\left(\Xi(u)\right) \sim - \frac{\Gamma(1 - \mu)}{4^{\mu} \Gamma(1 + \mu)} u^{2 \mu } \;,
\end{equation}
so that for $N \to \infty$, $P_{N}(\rho)$ in Eq. (\ref{rw}) becomes
\begin{equation}
P_{N}(\rho) = \frac{1}{N} S\Big(x = \frac{\rho}{N^{1/2\mu}}, \alpha = 2 \mu, \sigma = \frac{\Gamma(1 - \mu)}{\Gamma(1+\mu)}\Big),
\end{equation}
where $S(\cdot)$ is the one-sided, $2 \mu$-dimensional stable law with
index $\alpha = 2 \mu$ and scale $\sigma = \Gamma(1 -
\mu)/\Gamma(1+\mu)$ \cite{kol}. In consequence, the distribution
$P(\omega)$ is given, for $N \to \infty$, by
\begin{equation}
\label{onesided}
P(\omega) \sim \frac{(1 - \omega)^{\mu - 1}}{N \omega^{\mu + 1}} \, S\Big(x = \frac{1}{N^{1/2\mu}} \sqrt{\frac{1 - \omega}{\omega}}, \alpha = 2 \mu, \sigma = \frac{\Gamma(1 - \mu)}{\Gamma(1+\mu)}\Big).
\end{equation}
Asymptotic behavior of $S(\cdot)$ has been discussed in detail in
Ref. \cite{kol}. When $x \ll 1$, $S(\cdot) \to const$, and hence,
\begin{equation}
P(\omega) \sim \frac{(1 - \omega)^{\mu - 1}}{N},
\end{equation}
\ie, $P(\omega)$ diverges as $\omega \to 1$. On the other hand, when
$x \gg 1$, \ie, when $\omega$ is sufficiently close to $0$, $S(\cdot) \sim
1/x^{4 \mu}$ \cite{kol}, which yields
\begin{equation}
P(\omega) \sim N \omega^{\mu - 1}.
\end{equation}
This means that $P(\omega)$ diverges when $\omega \to 0$ and,
generally, in this domain $0 < \mu < 1$ the distribution has a
characteristic $U$-shaped form strongly skewed towards small values of
$\omega$ since the amplitudes differ by a factor $N^2$. Note that for $\mu = 1/2$ the
distribution in Eq.~(\ref{onesided}) becomes the Cauchy distribution so that the simple form in Eq.~(\ref{mu12}) follows immediately.

\section{The distribution $P(\omega)$ for exponentially-truncated first passage
time distributions. Two non-communicating searchers}

We turn next to the analysis of the distribution $P(\omega)$ in case
when the parent first passage time distribution $\Psi(\tau)$,
Eq.~(\ref{truncated_dist}), possesses moments of arbitrary order.

\begin{figure}[t]
  \centerline{\includegraphics*[width=0.65\textwidth]{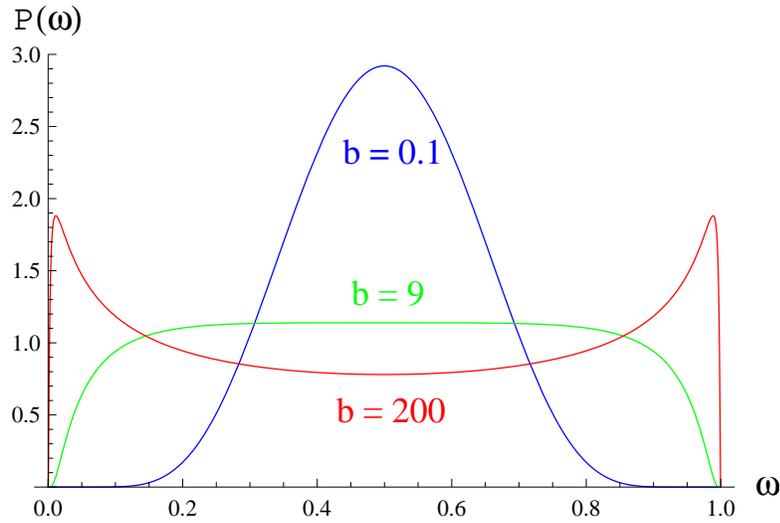}}
  \caption{The distribution $P(\omega)$ in Eq.~(\ref{2trunc}) for
    $\mu = 1/2$ and different values of $b$ ($a$ is set equal to $1$).}
  \label{fig5}
\end{figure}

For two non-communicating searchers and the exponentially-truncated
distribution in Eq.~(\ref{truncated_dist}), we get the following
result for the distribution of the random variable $\omega$:
\begin{equation}
\label{2trunc}
P(\omega) = \frac{1}{2 K^2_{\mu}(2 \sqrt{a/b})} \, \frac{1}{\omega (1 - \omega)} \, K_{2 \mu}\left(2 \sqrt{\frac{a}{b \, \omega ( 1- \omega)}}\right).
\end{equation}
Two remarks are in order. First, one readily notices that $P(\omega)$
vanishes exponentially fast when $\omega \to 0$ or $\omega \to 1$ so
that $P(\omega = 0) = P(\omega = 1) = 0$. Second, $P(\omega)$ is
clearly symmetric under the replacement $\omega \to 1 - \omega$. Since
here we deal with a truncated distribution $\Psi(\tau)$ which
possesses the moments of arbitrary order, our first guess would be
that $P(\omega)$ is always a bell-shaped function with a maximum at
$\omega = 1/2$. To check this guess, we expand $P(\omega)$ in the
Taylor series around $\omega = 1/2$:
\begin{eqnarray}
\fl P(\omega) = \frac{2 K_{2 \mu}(4 \sqrt{a/b})}{K^2_{\mu}(2 \sqrt{a/b})} \, \Big[ 1 + \nonumber\\
+
4 \left(1 - \mu - 2 \sqrt{\frac{a}{b}} \frac{K_{2 \mu - 1}(4 \sqrt{a/b})}{K_{2 \mu}(4 \sqrt{a/b})}\right) \left(\omega - \frac{1}{2}\right)^2 + \mathcal{O}\left(\omega - \frac{1}{2}\right)^4\Big].
\end{eqnarray}
Inspecting the sign of the coefficient before the quadratic term, \ie,
\begin{equation}
\label{g}
g = 1 - \mu - 2 \sqrt{\frac{a}{b}} \frac{K_{2 \mu - 1}(4 \sqrt{a/b})}{K_{2 \mu}(4 \sqrt{a/b})},
\end{equation}
we notice that
\begin{itemize}
\item For $\mu > 1$, $g$ is always negative for any value of $b/a$ so
  that here the distribution $P(\omega)$ is a bell-shaped function
  with a maximum at $\omega = 1/2$.
\item For $\mu = 1$, $g$ is negative and approaches $0$ from below
  when $b/a \to \infty$. It means that $P(\omega)$ is generally a
  bell-shaped function with a maximum at $\omega = 1/2$, but it is
  becoming progressively flatter when $b/a$ is increased, so that
  ultimately $P(\omega) \approx 1$ apart of very narrow regions at the
  edges for $b/a \gg 1$.
\item For $0 \leq \mu < 1$ there always exists a critical value
  $y_c(\mu)$ of the parameter $y = b/a$ which is defined implicitly as
  the solution of the equation $g = 0$, Eq.~(\ref{g}).  For $b/a <
  y_c(\mu)$, the distribution $P(\omega)$ is unimodal with a maximum
  at $\omega = 1/2$. For $b/a = y_c(\mu)$, the distribution is nearly
  uniform except for narrow regions in the vicinity of the
  edges. Finally, which is quite surprising in view of the fact that
  in this case $\Psi(\tau)$ possesses all moments, for $b/a >
  y_c(\mu)$ the distribution $P(\omega)$ is bimodal with a
  characteristic $M$-shaped form, two maxima close to $0$ and $1$ and
  $\omega = 1/2$ being the least probable value.
\end{itemize}
We depict in
  Fig.~\ref{fig5} three characteristic forms of $P(\omega)$ for $\mu =
  1/2$ and three different values of $b/a$.

Therefore, for $\mu \geq 1$ two non-communicating searchers will most
probably reach the target simultaneously. For $0 \leq \mu < 1$ two
distinctly different situations are possible: if $b/a$ is less than
some well-defined critical value $y_c(\mu)$, then most likely both
searchers will arrive to the target for the first time together.  If,
on the contrary, $b/a$ exceeds this critical value, the event in which both
searchers arrive to the location of the target simultaneously will be the
least probable one.

\subsection{Two independent BMs in a semi-infinite 1D system with a bias}

Recall now that the distribution in Eq.~(\ref{truncated_dist}) with
$\mu = 1/2$ is exact for a BM in semi-infinite 1D systems
in presence of a constant bias pointing towards the target. In this
case one has $a = x_0^2/4 D$, where $x_0$ is the starting point and
$D$ - the diffusion coefficient, and $b = 4 D/v^2$, $v$ being the
drift velocity. Hence, $2 \sqrt{a/b} = {\it Pe} = x_0 |v|/2 D$ is the
Peclet number (see, \eg, Ref.~\cite{redner}). Consequently, we can make
a following statement:

Consider two independent, absolutely identical BMs on a semi-infinite
line, starting at the same point $x_0$, having the same diffusion
coefficient $D$ and experiencing the same bias $F$ which points towards
the origin so that the drift velocity of both BMs is $v < 0$. Then, an event in
which both BMs arrive for the first time to the origin simultaneously
is
\begin{itemize}
\item the \textit{least} probable if $Pe < Pe_c$\,,
\item the \textit{most} probable if $Pe > Pe_c$\,,
\end{itemize}
where $Pe_c$ is the solution of the transcendental equation
\begin{equation}
1 = 2 Pe_c \frac{K_0(2 Pe_c)}{K_1(2 Pe_c)}.
\end{equation}
An approximate solution of the latter equation gives $Pe_c \approx
0.666...$.

Therefore, the mean first passage time to the target might be an
appropriate measure of the search efficiency for sufficiently large
Peclet numbers, but definitely is not the one in case of small $Pe$. In
the latter case the sample-to-sample fluctuations are significant and
the mean value is not representative of the actual behaviour.

\subsection{Two independent unbiased BMs on a finite interval}

We have already remarked that the expression in
Eq.~(\ref{truncated_dist}) is a reasonable approximation for the first
passage time distribution for random motion in finite systems.  For a
one-dimensional bounded interval of length $L$ and for $\mu = 1/2$,
the parameter $b \sim L^2/D$, while $a \sim x_0^2/D$, where $x_0$ is
the starting point.  Consequently, $\sqrt{a/b}$ should be $ \sim
x_0/L$ and independent of the particles' diffusion coefficient $D$.
This suggests a somewhat strange result that the very shape (or the
modality) of the distribution $P(\omega)$ will crucially depend on the
starting point $x_0$.

\begin{figure}[ht]
  \centerline{\includegraphics*[width=0.65\textwidth]{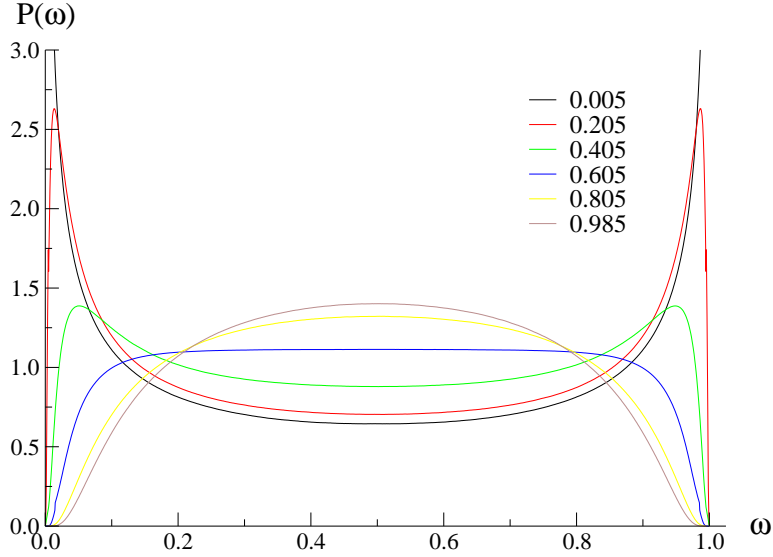}}
  \caption{Two BMs on a finite interval $[0,L]$ with a
    reflecting boundary at $x = L$.  The distribution $P(\omega)$ in
    Eq.~(\ref{pw1d}) for different values of $x_0/L$, $x_0$ being the
    starting point of both BMs.}
  \label{1dd}
\end{figure}

To verify such a prediction, we focus now on a BM on a finite
1D interval and re-examine this question using an exact form of the
normalized first passage time distribution $\Psi(\tau)$.  For the BM
with a starting point at $x_0$ on an interval of length $L$ with an
adsorbing boundary at $x=0$ and a reflecting boundary at $x = L$,
$\Psi(\tau)$ is given by
\begin{equation}
\label{fpt1}
\Psi(\tau) = \frac{2 \pi D}{L^2} \sum_{n = 0}^{\infty} A^{(d=1)}_n\left(\frac{x_0}{L}\right) \exp\left(- \frac{\pi^2 (n + 1/2)^2 D \tau}{L^2}\right),
\end{equation}
where
\begin{equation}
A^{(d=1)}_n\left(\frac{x_0}{L}\right) = \left(n + \frac{1}{2}\right)\sin\left(\frac{\pi (n + 1/2) x_0}{L}\right) \ .
\end{equation}
Consequently, the normalized distribution $P(\omega)$ in case of two
independent, identical BMs has the following form
\begin{eqnarray}
\label{pw1ddd}
P(\omega) &=& \frac{4}{\pi^2} \sum_{m,n=0}^{\infty} \frac{A^{(d=1)}_n\left(x_0/L\right) \, A^{(d=1)}_m\left(x_0/L\right)}{\left(\omega (n + 1/2)^2 + (1 - \omega) (m+ 1/2)^2\right)^2}.
\end{eqnarray}
Using next an equality
\begin{equation}
\label{equality}
\frac{1}{\lambda_m^2} \frac{d}{d\omega} \frac{1}{\lambda_n^2 + \frac{1 - \omega}{\omega} \lambda_m^2} = \frac{1}{\left(\omega \lambda_n^2 + (1 - \omega) \lambda_m^2\right)^2},
\end{equation}
and the following representation of the characteristic function $\Phi(\lambda)$ of the
distribution $\Psi(\tau)$ in Eq.~(\ref{fpt1}),
\begin{eqnarray}
\Phi(\lambda) &=& \int_0^{\infty} d\tau \, \Psi(\tau) \, \exp\left( - \lambda \, \tau\right)  \nonumber\\
&=& \frac{2}{\pi} \sum_{n = 0}^{\infty} \frac{ A^{(d=1)}_n\left(x_0/L\right)}{(n + 1/2)^2 + L^2 \lambda/\pi^2  D}  \nonumber\\
&=& \frac{\cosh\left((L - x_0) \sqrt{\lambda/D}\right)}{\cosh\left(L \sqrt{\lambda/D}\right)},
\end{eqnarray}
we can conveniently rewrite Eq.~(\ref{pw1ddd}) as
\begin{eqnarray}
\label{pw1d}
P(\omega) = \frac{2}{\pi} \frac{d}{d\omega} \sum_{m = 0}^{\infty} \frac{\sin\left(\pi (m+1/2) \frac{x_0}{L}\right)}{m + 1/2} \, \Phi\left(\lambda = \frac{\pi^2 D}{L^2} \frac{1 - \omega}{\omega} \left(m + 1/2\right)^2\right)  \nonumber\\
= \frac{2}{\pi} \frac{d}{d\omega} \sum_{m = 0}^{\infty} \frac{\sin\left(\pi (m+1/2) \frac{x_0}{L}\right)}{m + 1/2} \, \frac{\cosh\left(\pi (m+1/2) \sqrt{\frac{1 - \omega}{\omega}} \left(1 - \frac{x_0}{L}\right)\right)}{\cosh\left(\pi (m+1/2) \sqrt{\frac{1 - \omega}{\omega}}\right)} \,.
\end{eqnarray}

 The distribution $P(\omega)$ tends to zero exponentially, $P(\omega) \sim \exp(- \pi x_0/2 L \sqrt{\omega})/\omega^{3/2}$, when $\omega \to 0$, precisely in the same way as $P(\omega)$ in Eq.~(\ref{2trunc}) obtained
  for the exponentially truncated first passage time distribution.
  By symmetry, we expect the same behavior when $\omega \to 1$. In principle, $P(\omega)$ in
Eq.~(\ref{pw1d}) can be represented in closed form as a complicated
combination of elliptic function. However,
we prefer to proceed with a numerical analysis of the rapidly convergent series in  Eq.~(\ref{pw1d}),
 in order to understand whether $P(\omega)$ is always a bell-shaped function of $\omega$, or undergoes a transition to an $M$-shaped form at a certain value of $x_0/L$.

In Fig.~\ref{1dd} we depict the distribution $P(\omega)$ in
Eq.~(\ref{pw1d}) for different values of the ratio $x_0/L$. One
notices that $P(\omega)$ has a different shape depending whether
$x_0/L$ is less or greater than the critical value $\approx 0.605...$. This allows us to make the following statement:

Consider two independent, identical, unbiased BMs starting at the same
point $x_0$ on a finite interval $[0,L]$ with a reflecting boundary
 at $x = L$. Then, an event in which both BMs arrive
simultaneously to the origin is
\begin{itemize}
\item the \textit{least} probable if $x_0/L \lesssim 0.605...$,
\item the \textit{most} probable if $x_0/L \gtrsim 0.605...$.
\end{itemize}

From a common sense point of view such a behavior seems a bit
counterintuitive - indeed, why should two BMs arrive to the target at
progressively distinct times the closer they are to its location, and
should most probably arrive together when they are far from it?
On
the other hand,  such a behavior
is quite a natural one:  Indeed, $x_0$ ($a^{1/2}$) and
$L$ ($b^{1/2}$) define the effective size of the window in which the
decay of the first passage time distribution is governed by the
intermediate power-law tail. The larger is this window, the closer we
are to the situation described in Section III.
Therefore, the origin
of such a disproportionate behavior of two identical, independent BMs
is precisely the same as the one behind the famous arcsine law for the
distribution of the fraction of time spent by a random walker on a
positive half-axis \cite{arcsine}: Once one of the BMs goes away from
the target, it finds it more difficult to return than to keep on going
away.

\begin{figure}[t]
  \centerline{\includegraphics*[width=0.7\textwidth]{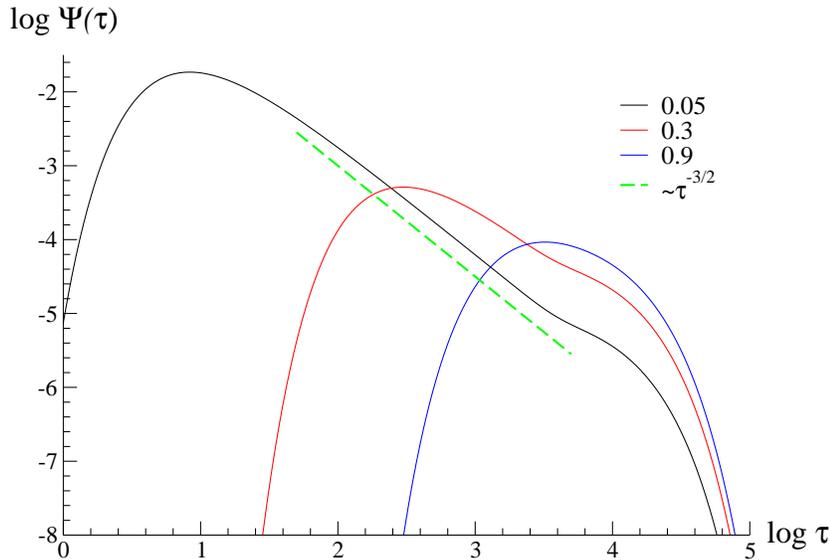}}
  \caption{A BM on a finite interval $[0,L]$ (with a
    reflecting boundary at $x = L$) starting at $x = x_0$. The first
    passage time distribution $\Psi(\tau)$ in Eq.~(\ref{fpt1}) for $L=
    100$, $D = 1/2$ and different values of $x_0/L$.}
  \label{fpt1dd}
\end{figure}

To substantiate this claim, we plot in Fig.~\ref{fpt1dd} the
distribution in Eq.~(\ref{fpt1}) for three different values of $x_0$
and fixed $L$. Note that all three curves show an exponential behavior
for both small and large values of $\tau$ (which, in fact, is an argument in favor of our choice
of the exponentially truncated distribution in Eq.~(\ref{truncated_dist})). For large $\tau$ all three
curves merge which signifies that at such values of $\tau$ the
characteristic decay time is dependent only on $L$. The lower cut-off
is clearly dependent only on the starting point $x_0$. Further on,
notice that the closer (for a fixed $L$) the starting point $x_0$ to
the reflecting boundary is, the narrower is the distribution
$\Psi(\tau)$. On contrary, the smaller $x_0$ is (the closer to the
target), the more pronounced the intermediate power-law behavior $\sim
t^{-3/2}$ becomes (see the dashed line in Fig.~\ref{fpt1dd}). Actually, the fact that the farther away the starting point
from the reflecting wall is,
the broader is the first passage time distribution (the BM simply does not "know" that it is in a finite system up to times of order $\sim L^2/D$) has been already discussed in detail in Ref.~\cite{redner}.

\begin{figure}[ht]
  \centerline{\includegraphics*[width=0.65\textwidth]{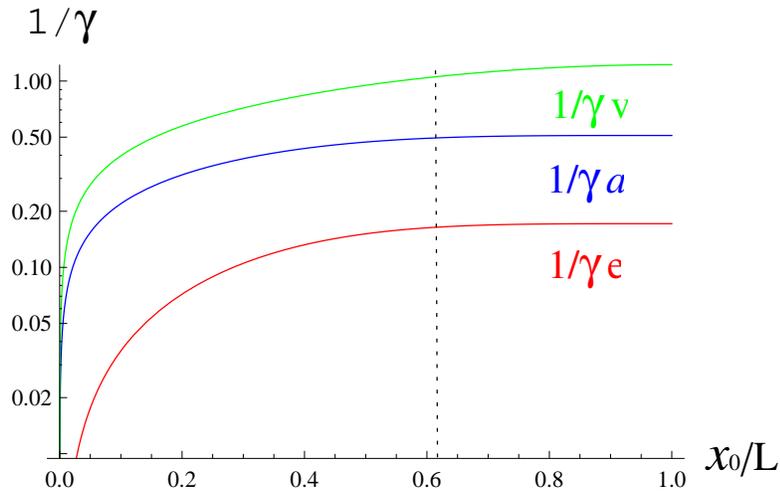}}
\caption{The coefficient of variation $\gamma_v$, the skewness $\gamma_a$ and the kurtosis $\gamma_e$
of the
distribution $\Psi(\tau)$ in Eq.~(\ref{fpt1}) versus $x_0/L$. Dotted vertical line defines the critical value
$x_0/L \approx 0.605...$ at which the distribution $P(\omega)$ changes the modality.}
  \label{kurtosis}
\end{figure}

Consider next some standard characteristics of the first-passage-time distribution
in Eq.~(\ref{fpt1}), such as the mean first passage time, the skewness, the coefficient of variation and the kurtosis,
to see if any of them reflects the transition observed for $P(\omega)$ at $x_0/L \approx 0.605...$.
Note that the moments of arbitrary order of the distribution $\Psi(\tau)$ in Eq.~(\ref{fpt1}) can be straightforwardly calculated in closed form:
\begin{equation}
\big<\tau^m\big> \equiv \int^{\infty}_0 \tau^m \, d\tau \, \Psi(\tau) = \frac{(-1)^m \sqrt{\pi}}{\Gamma(m + 1/2)} \left(\frac{ L^2}{D}\right)^m E_{2 m}\left(\frac{x_0}{2 L}\right) \,,
\end{equation}
where $E_{2 k}(.)$ are the Euler polynomials. Consequently, the mean and the variance are given by
\begin{equation}
\big<\tau\big> = - 2 \frac{ L^2}{D} \, E_2\left(\frac{x_0}{2 L}\right),
\end{equation}
and
\begin{equation}
k_2 = \big<\tau^2\big> - \big<\tau\big>^2 = \frac{4}{3} \frac{L^4}{D^2} \left(E_4\left(\frac{x_0}{2 L}\right) - 3 E_2^2\left(\frac{x_0}{2 L}\right)\right) \,
\end{equation}
respectively. One can readily check that both characteristic properties are
monotonically increasing functions of $x_0/L$ and do not show any sign of a particular behavior at
$x_0/L \approx 0.605...$. Further on, we define the coefficient of variation
\begin{equation}
\gamma_v = \sqrt{\frac{k_2^2}{\big<\tau\big>^2}} = \sqrt{\frac{E_4\left(x_0/2 L\right) - 3 E_2^2\left(x_0/2 L\right)}{3 E_2^2\left(x_0/2 L\right)}},
\end{equation}
the skewness
\begin{equation}
\gamma_a = \frac{k_3}{k_2^{3/2}},
\end{equation}
where $k_3$ is the third cumulant:
\begin{eqnarray}
k_3 = \big<\tau^3\big> - 3 \big<\tau\big> \big<\tau^2\big> + 2 \big<\tau\big>^3,
\end{eqnarray}
and the kurtosis (coefficient of excess)
\begin{equation}
\gamma_e = \frac{k_4}{k_2^2},
\end{equation}
with $k_4$ being the fourth cumulant of the distribution in Eq.~(\ref{fpt1}):
\begin{equation}
k_4 = \big<\tau^4\big> - 3 \big<\tau^2\big>^2 - 4 \big<\tau\big> \big<\tau^3\big> + 12 \big<\tau\big>^2 \big<\tau^2\big> - 6\big<\tau\big>^4 \,.
\end{equation}
In Fig.~\ref{kurtosis} we plot the coefficient of variation, the skewness and the kurtosis of the distribution in Eq.~(\ref{fpt1}). One observes a strong variation of these properties reflecting
an influence of extreme events for sufficiently small values of $x_0/L$. For $x_0/L > 0.6$ the variation of $\gamma_a$ and $\gamma_e$ becomes rather small.
However, neither of these properties
shows a clear demarkation line between different regimes exhibited by $P(\omega)$.

We finally remark that in  many  practically interesting physical problems
the  starting
point $x_0$ is not fixed,
but the searcher rather starts from some random
location which is uniformly distributed on the interval.
The distribution $P_{av}(\omega)$ appropriate to such a situation
is obtained by merely
averaging $P(\omega)$ in Eq.~(\ref{pw1d})
over $x_0$, \ie,
\begin{equation}
\label{av}
P_{av}(\omega) = \frac{1}{L} \int^L_{0} dx_0 \, P(\omega).
\end{equation}
A quick inspection of $P(\omega)$ in Eq.~(\ref{pw1d}) shows that,
due to the orthogonality of $A^{(d=1)}_n(x_0/L)$, the averaged distribution $P_{av}(\omega)$
does not depend on $\omega$, so that
\begin{equation}
P_{av}(\omega) \equiv 1.
\end{equation}
This seems to be  a general property of $P_{av}(\omega)$, associated with
the probability conservation and thus, as will be
checked for further examples, is independent of the dimension of space.

\subsection{Two independent BMs in a disc with a reflecting boundary}

Recall that in an infinite 2D system the first passage
time distribution has an algebraic (with a logarithmic correction)
tail with $\mu = 0$, Eq.~(\ref{distribution_log}). Consequently, we
may expect essentially the same behavior as we observed
in two previous subsections.

\begin{figure}[ht]
  \centerline{\includegraphics*[width=0.45\textwidth]{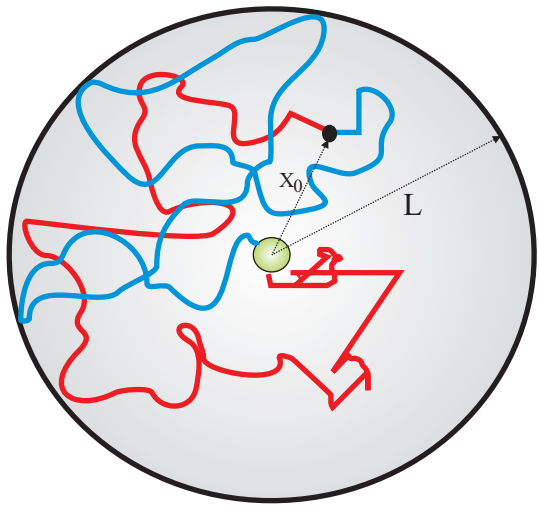}}
  \caption{Two BMs in a (2D or 3D) bounded spherical domain
  of radius $L$ with a reflecting (thick black line) boundary. The target is a circle (in 3D, a sphere) of radius $r$ centered at the origin. Two BMs start from the same point (a filled circle) at a distance $x_0$ from the center and arrive to the target boundary for the first time at time moments $\tau_1$ and $\tau_2$, respectively.}
  \label{target2}
\end{figure}

Consider an immobile target of radius $r$ fixed at the origin of a
disc of radius $L$ with a reflecting boundary. Suppose next that a BM
starts at some point at distance $x_0$ from the origin and hits the
target for the first time at time moment $\tau$. Then, the first
passage time distribution $\Psi(\tau)$ in such a situation is given
explicitly by

\begin{figure}[t]
  \centerline{\includegraphics*[width=0.7\textwidth]{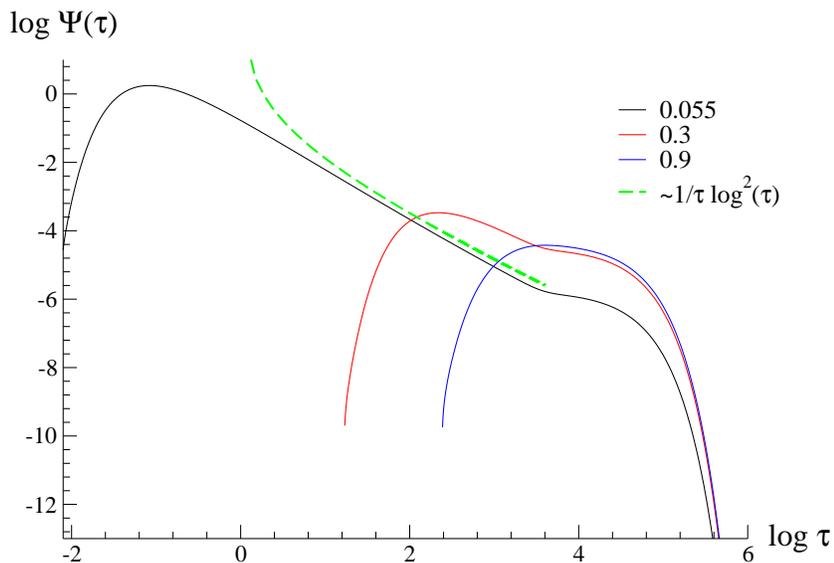}}
  \caption{A BM in a two-dimensional disc of radius $L$ with a reflecting boundary.
First passage
    time distribution $\Psi(\tau)$ in Eq.~(\ref{fpt2}) for $D = 1/2$,
    $L= 100$, $r = 5$ and different values of $x_0/L$.}
  \label{fpt2d}
\end{figure}

\begin{equation}
\label{fpt2}
\Psi(\tau) = \frac{r D}{Z} \sum_{n = 0}^{\infty} A_n^{(d = 2)}(r,x_0,L)  \exp\left(- \lambda_n^2 D \tau\right),
\end{equation}
where $Z$ is the normalization,
\begin{equation}
A_n^{(d = 2)}(r,x_0,L) = \frac{U_0(\lambda_n x_0) U_0'(\lambda_n r)}{\lambda_n^2 L^2 U_0^2(\lambda_n L) - r^2 \left(U_0'(\lambda_n r)\right)^2},
\end{equation}
\begin{equation}
U_0(x) = Y_0(\lambda_n r) J_0(x) - J_0(\lambda_n r) Y_0(x),
\end{equation}
\begin{equation}
U_0'(\lambda_n r) = \frac{d U_0(\lambda_n x)}{dx}\bigg|_{x=r},
\end{equation}
while $\lambda_n$ are the roots of the function
\begin{equation}
Y_0(\lambda_n r) J_1(\lambda_n L) - J_0(\lambda_n r) Y_1(\lambda_n L),
\end{equation}
arranged in an ascending order, and $Y_n(\cdot)$ are Bessel functions
of the second kind. Note that $\lambda_n$ depends on
$L$ and $r$.

In Fig.~\ref{fpt2d} we depict the distribution in Eq.~(\ref{fpt2}) for
fixed $L$, $D$ and $r$, and several values of $x_0$.
Note that, apart of the intermediate behavior, which follows a
power-law with a logarithmic correction $\sim 1/\tau \log^2(\tau)$ (see
the dashed line in Fig.~\ref{fpt2d}), we have here essentially the
same trend as the one we observed for a BM on a 1D finite
interval. Namely, the closer is the starting point to the location of
the target, the broader is the first passage time distribution so that
the intermediate power-law behavior gets more pronounced.

Suppose next that we have two BMs starting at the same point some
distance $x_0$ apart of the origin. Then, from Eq.~(\ref{fpt2}), we
get the following result for the distribution $P(\omega)$:
\begin{equation}
\label{pw222}
P(\omega) = \frac{r^2}{Z^2} \sum_{n,m=0}^{\infty} \frac{A_n^{(d=2)}(r,x_0,L) \, A_m^{(d = 2)}(r,x_0,L)}{\left(\omega \lambda_n^2 + (1 - \omega) \lambda_m^2\right)^2} \,,
\end{equation}
which can be conveniently rewritten, using the equality in Eq.~(\ref{equality}), as
\begin{equation}
 \label{pw2}
P(\omega) = \frac{r}{Z} \frac{d}{d\omega} \sum_{m=0}^{\infty} \frac{A_m^{(d = 2)}(r,x_0,L)}{\lambda_m^2} \, \Phi\left(\lambda = \frac{1 - \omega}{\omega} D \lambda_m^2\right) \,,
 \end{equation}
where $\Phi(\lambda)$ is the characteristic function of
the first passage time distribution in Eq.~(\ref{fpt2}) defined by \cite{redner}:
\begin{equation}
\Phi(\lambda) = \frac{I_0\left(\sqrt{\frac{\lambda}{D}} x_0\right) K_1\left(\sqrt{\frac{\lambda}{D}} L\right)   + K_0\left(\sqrt{\frac{\lambda}{D}} x_0\right) I_1\left(\sqrt{\frac{\lambda}{D}} L\right)}{I_0\left(\sqrt{\frac{\lambda}{D}} r\right) K_1\left(\sqrt{\frac{\lambda}{D}} L\right)   + K_0\left(\sqrt{\frac{\lambda}{D}} r\right) I_1\left(\sqrt{\frac{\lambda}{D}} L\right)} \,.
\end{equation}

We depict the result in Eq.~(\ref{pw2}) for fixed $L$ and $r$, and
several values of $x_0/L$.
\begin{figure}[t]
  \centerline{\includegraphics*[width=0.65\textwidth]{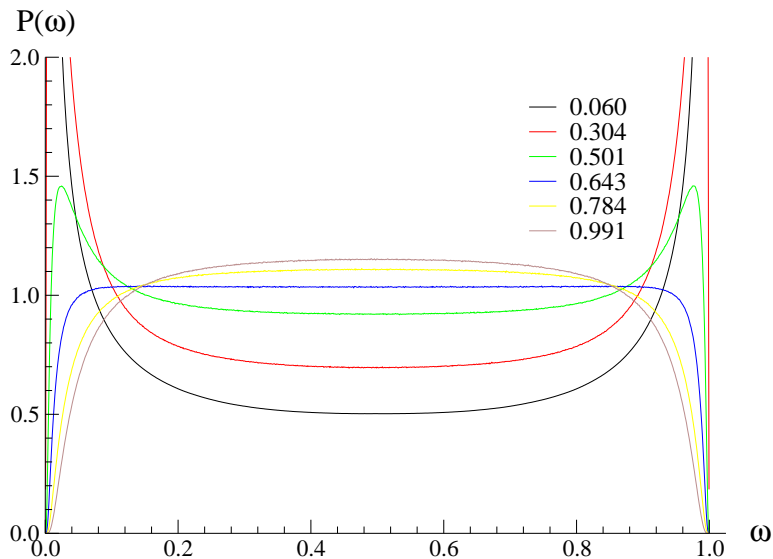}}
  \caption{Two Brownian motions in a two-dimensional disc. The
    distribution $P(\omega)$ in Eq.~(\ref{pw2}) for $L= 100$, $r = 5$
    and different values of $x_0/L$.}
  \label{pw2d}
\end{figure}
One notices that the shape of the distribution $P(\omega)$ is again
different, depending on the value of the ratio $x_0/L$. For $x_0/L
\lesssim 0.643$, the distribution has an M-shaped form with a minimum
at $\omega = 1/2$, for $x_0/L \approx 0.643$ the distribution is
nearly uniform except for narrow regions near the edges, and lastly,
for $x_0/L \gtrsim 0.643$ it becomes unimodal, but still it is very
broad and has a pronounced flat region around the maximum. This
signifies that even in this regime the sample-to-sample fluctuations
are significant.



Consider finally the situation when the starting point of two BM is uniformly distributed
within the space between two concentric circles of radius $r$
 and radius $L$. One readily finds, by averaging the
result in Eq.~(\ref{pw2}), that here as in 1D one has $P_{av}(\omega) \equiv 1$.

\subsection{Two independent BMs in a sphere with a reflecting boundary}

We  now turn  our attention  to a Brownian  motion in  a  3D spherical domain with a reflecting
boundary.
The target is supposed to be a sphere of
radius $r$ which is fixed at the origin.

The distribution of the first  passage time $\tau$ of a BM
starting at a distance $x_0$ from the origin to the surface of the target is given explicitly by
\begin{equation}
\label{fpt3}
\Psi(\tau) = \frac{D}{Z} \sum_{n=0}^\infty A^{(d=3)}_n(r,x_0,L)
\exp\left(-\lambda_n^2D\tau\right) \ ,
\end{equation}
where
\begin{equation}
A^{(d=3)}_n(r,x_0,L) = \frac{2 u_0(\lambda_n x_0) u_0^\prime(\lambda_n r)}
{G(r,L,\lambda_n)} \ ,
\end{equation}
\begin{equation}
u_0(x) = y_0(\lambda_n r) j_0(x) - j_0(\lambda_n r) y_0(x) \ ,
\end{equation}
the derivative $u_0^\prime(\lambda_n r) = du_0(\lambda_n x)/dx|_{x=r}$, and
\begin{eqnarray}
G(r,L,\lambda_n) &=& Lj_0(\lambda_nL)
u_0(\lambda_nL)(\lambda_nLj_0(\lambda_nr) - y_0(\lambda_nr))
- \nonumber\\
&-& \frac{j_0(\lambda_nr)y_0(\lambda_nr)}{\lambda_n}
 + \frac{L-r}{r^2\lambda_n^4} \ .
\end{eqnarray}
The set $\{\lambda_n\}$ are the roots of $ y_0(\lambda_n r)
j_1(\lambda_n L) - j_0(\lambda_n r) y_1(\lambda_n L) $, arranged in
an ascending order, while $j_n(\cdot)$ and $y_n(\cdot)$ are the spherical Bessel
functions of the first and of the second kind, respectively.

\begin{figure}[t]
  \centerline{\includegraphics*[width=0.7\textwidth]{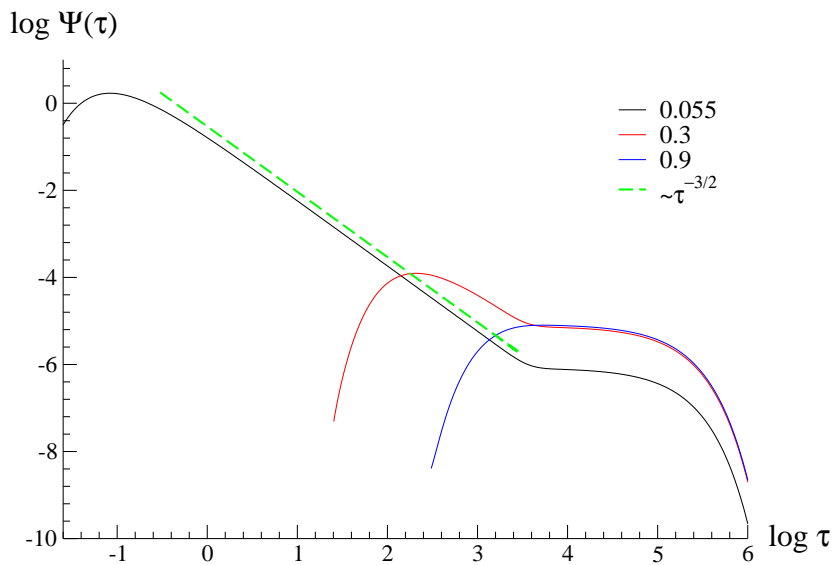}}
  \caption{A BM in a sphere with a reflecting boundary. First passage time distribution $\Psi(\tau)$ in Eq.~(\ref{fpt3}) for
$L= 100$, $r=5$, $D = 1/2$ and different
    values of $x_0/L$.  }
  \label{fpt3d}
\end{figure}

The distribution of the first  passage time in Eq.~(\ref{fpt2}) is plotted in
Fig.~\ref{fpt3d}   for  three  different   values  of   $x_0/L$.   An
intermediate  power-law $\sim  \tau^{-3/2}$  is apparent  for
$x_0/L=0.055$, persists for about a decade for $x_0/L=0.3$ and is entirely absent for $x_0/L=0.9$, \ie, when
the BM starts close to the reflecting boundary.

From Eq.~(\ref{fpt3}), we obtain
the normalized distribution $P(\omega)$ for two
independent BMs starting at a distance $x_0$ from the origin:
\begin{eqnarray}
\label{pw3}
P(\omega) &=& \frac{1}{Z^2} \sum_{n,m=0}^\infty
\frac{A^{(d=3)}_n(r,x_0,L) A^{(d=3)}_m(r,x_0,L)}{\left(\omega \lambda_n^2 + (1 - \omega) \lambda_m^2\right)^2} = \nonumber\\
&=& \frac{1}{Z} \sum_{m=0}^\infty \frac{A^{(d=3)}_m(r,x_0,L)}{\lambda_m^2} \, \Phi\left(\lambda = \frac{1 - \omega}{\omega} D \lambda_m^2\right) \,,
\end{eqnarray}
where $\Phi(\lambda)$ is the characteristic function of $\Psi(\tau)$ in Eq.~(\ref{fpt3}), defined by \cite{redner}
\begin{equation}
\Phi(\lambda) = \frac{r}{x_0} \frac{\sinh\left(\sqrt{\frac{\lambda}{D}} \left(L - x_0\right)\right) - \sqrt{\frac{\lambda}{D}} \, L \, \cosh\left(\sqrt{\frac{\lambda}{D}} \left(L - x_0\right)\right)}{ \sinh\left(\sqrt{\frac{\lambda}{D}} \left(L - r\right)\right) - \sqrt{\frac{\lambda}{D}} \, L \, \cosh\left(\sqrt{\frac{\lambda}{D}} \left(L - r\right)\right)} \,.
\end{equation}

\begin{figure}[t]
  \centerline{\includegraphics*[width=0.65\textwidth]{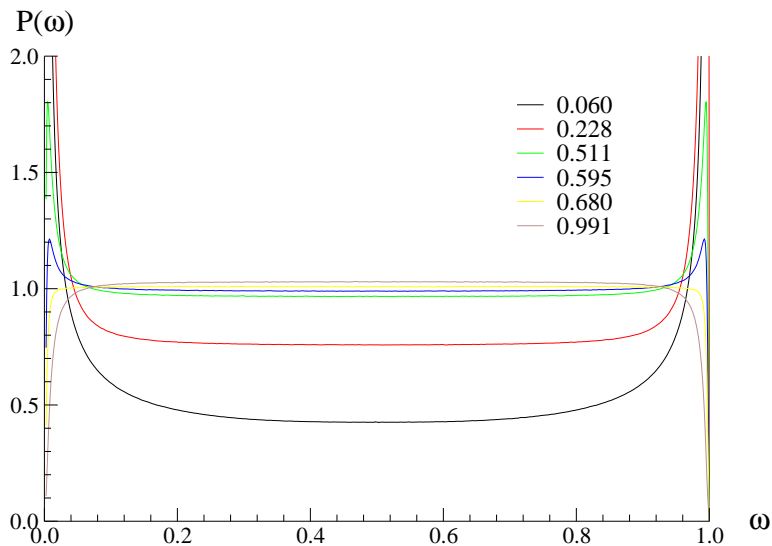}}
  \caption{Two BMs in a sphere with a reflecting boundary. The distribution $P(\omega)$ in Eq.~(\ref{pw3})
for $L= 100$, $r=5$ and different values of
    $x_0/L$.  }
  \label{pw3d}
\end{figure}

This distribution is plotted in Fig.~(\ref{pw3d}) for fixed $L$ and $r$, and several
values of $x_0/L$.
Therefore, also in 3D we find that $P(\omega)$ has a different modality depending on the value of the ratio $x_0/L$.
For $x_0/L \lesssim 0.68..$, the distribution has an M-shaped form with $\omega = 1/2$ being the least probable value.
For $x_0/L > 0.68..$, the distribution has, in principle, a maximum at $\omega = 1/2$
but this maximum is almost invisible so that visually the distribution looks more like a uniform one, as compared to the 1D case in which
the maximum is more apparent.

In case when the starting point $x_0$ is uniformly
distributed between two concentric spheres of radii $r$
and $L$, we again find $P_{av}(\omega) \equiv 1$.

\subsection{Two independent BMs in a sphere with an adsorbing boundary}

Consider finally a geometrically different situation in which two BMs
start from the same point within a three-dimensional sphere at a fixed
distance $x_0$ from the origin but now the target is the surface of
the sphere. In this case, the normalized distribution of the time of
the first passage of a BM with diffusion coefficient $D$ to any point
on the surface of the sphere of radius $L$ from a point at distance
$x_0$ from the origin is given by the series (see, \eg,
\cite{net1,gros}):
 \begin{equation}
 \label{ads1}
 \Psi(\tau) = \frac{2 \pi D}{x_0 L} \sum_{n=1}^{\infty} (-1)^{n + 1} n \, \sin\left(\pi n \frac{x_0}{L}\right) \, \exp\left(- \frac{\pi^2 n^2 D \tau}{L^2}\right),
 \end{equation}
whose moments of arbitrary order $m$ are defined as
\begin{equation}
\big<\tau^m\big> = \frac{(-1)^m \sqrt{\pi}}{\Gamma(m + 3/2)} \, \frac{L}{x_0} \left(\frac{L^2}{D}\right)^m \, B_{2 m + 1}\left(\frac{x_0 + L}{2 L}\right),
\end{equation}
where $B_m(.)$ are the Bernoulli polynomials.
\begin{figure}[t]
  \centerline{\includegraphics*[width=0.7\textwidth]{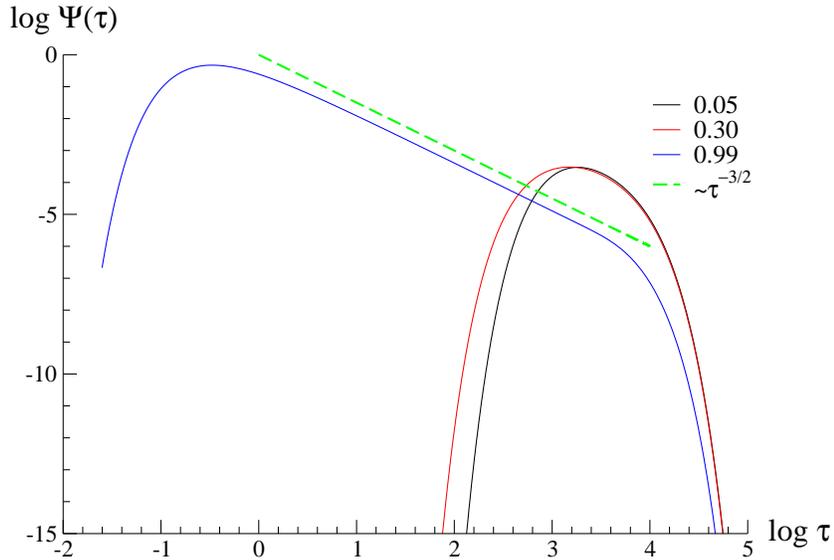}}
  \caption{A BM starting at a distance
  $x_0$ from the origin of a sphere with an adsorbing
   boundary. The first passage time distribution
   $\Psi(\tau)$ in Eq.~(\ref{ads1}) for $L=
    100$, $D = 1/2$ and different values of $x_0/L$.}
  \label{f1}
\end{figure}
The distribution in Eq.~(\ref{ads1}) is depicted in Fig.~\ref{f1} for fixed $L$ and $D$, and different values of the ratio $x_0/L$.
Note that here the situation is inverse to the one in which the target is situated in the origin -
the most pronounced intermediate time power-law behavior is observed for
$x_0/L \sim 1$ and is absent for small values of $x_0/L$.
\begin{figure}[t]
  \centerline{\includegraphics*[width=0.7\textwidth]{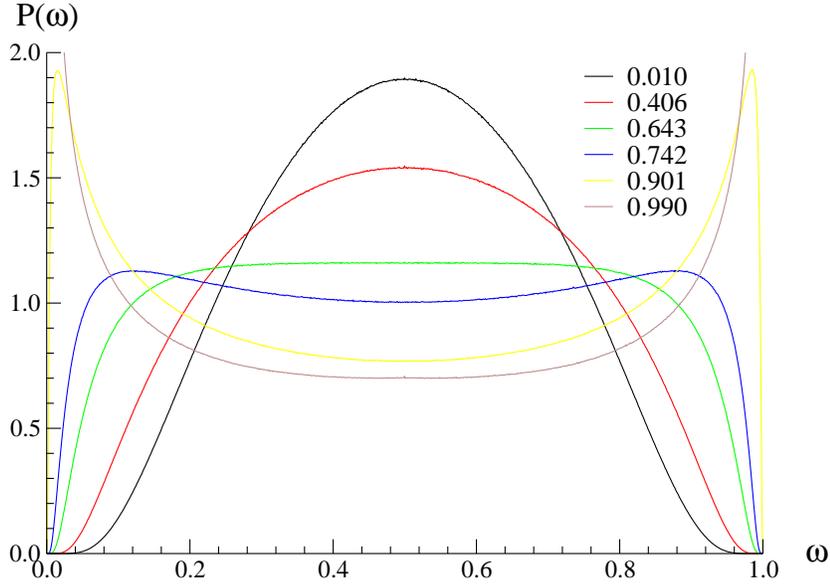}}
  \caption{Two BMs starting at a distance
  $x_0$ from the origin of a sphere with an adsorbing
   boundary. The distribution
   $P(\omega)$ in Eq.~(\ref{ads}) for $L=
    100$ and different values of $x_0/L$.}
  \label{f2}
\end{figure}
The normalized distribution $P(\omega)$ in this case has the form:
\begin{eqnarray}
\label{ads}
P(\omega)  &=& \frac{4 L^2}{\pi^2 x_0^2} \sum_{n,m=1}^{\infty} (-1)^{n + m} \frac{n \, m \, \sin\left(\pi n x_0/L\right) \,
\sin\left(\pi m x_0/L\right)}{\left(\omega n^2 + (1 - \omega) m^2\right)^2} \nonumber\\
&=& - \frac{i L^2}{2 \pi x_0^2} \, \frac{1}{\omega^{3/2} \left(1 - \omega\right)^{1/2}} \left. \frac{d}{d y} \ln\left(\frac{\theta_3\left(\frac{x_0 ( 1 + i y)}{2 L}, e^{- \pi y}\right)}{\theta_3\left(\frac{x_0 (1 - i y)}{2 L}, e^{- \pi y}\right)}\right)\right|_{y = \sqrt{\frac{1 - \omega}{\omega}}} \,
\end{eqnarray}
where $\theta_3$ is the Jacobi theta-function:
\begin{equation}
\theta_3(v,q) = \sum_{m=-\infty}^{\infty} q^{n^2} \exp\left(2 \pi i m v\right).
\end{equation}
The distribution in Eq.~(\ref{ads}) is depicted in Fig.~\ref{f2} for fixed $L$ and different values of $x_0/L$. We observe here
a transition from a bell-shaped form with a maximum at $\omega = 1/2$ and an M-shaped form with a minimum
at $\omega = 1/2$ and maxima close to $0$ and $1$. The transition takes place at $x_0/L = 0.643...$. Averaging Eq.~(\ref{ads}) over
the starting point $x_0$, we again find that $P_{av}(\omega) \equiv 1$.

\section{The distribution $P(\omega)$ for exponentially-truncated first passage time distributions.
 Three and more non-communicating searchers}

We turn finally to the situation with more than two searchers whose
first passage time distribution obeys an exponentially-truncated form
in Eq. (\ref{truncated_dist}). For $N = 3$ we find the following
general result
\begin{eqnarray}
\label{k}
P(\omega) &=& \frac{(b/4 a)^{3 \mu/2}}{2 K^3_{\mu}\left(2 \sqrt{a/b}\right)} \frac{\omega^{\mu - 1}}{(1 - \omega)^{\mu + 1}}
\int^{\infty}_0 x^{\mu} dx J_{\mu - 1}\left(x\right) \nonumber\\ &\times& \left(x^2 + \frac{4 a}{b \omega}\right)^{\mu}
 K^2_{\mu}\left(\sqrt{\frac{\omega}{1 - \omega}\left(x^2 + \frac{4 a}{b \omega}\right)} \right).
\end{eqnarray}
\begin{figure}[ht]
  \centerline{\includegraphics*[width=0.65\textwidth]{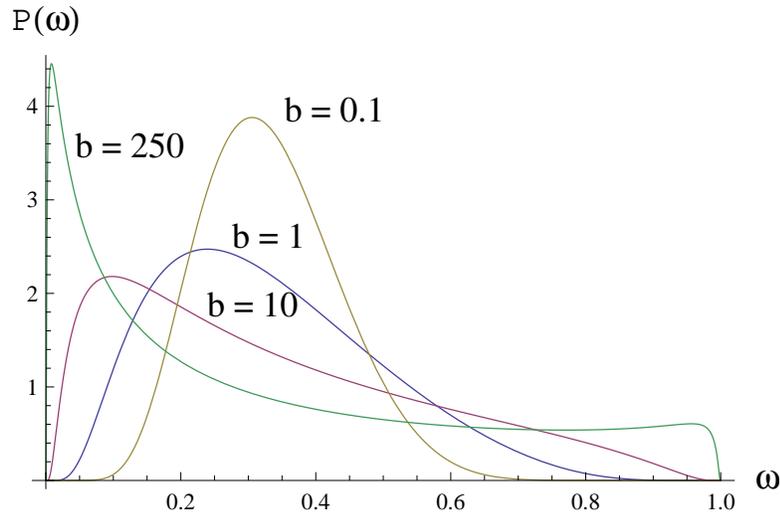}}
  \caption{Three non-communicating Brownian searchers. $P(\omega)$ in
    Eq. (\ref{3exp}) for different values of the parameter $b$ ($a$ is
    set equal to $1$).}
  \label{sketch6}
\end{figure}
A straightforward analysis shows that $P(\omega)$ in Eq. (\ref{k}) is
always a bell-shaped function for $\mu \geq 1$. The most probable
value $\omega_m$ is, however, always substantially less than $1/3$,
approaching this value only when $\mu \to \infty$ or $b \to 0$.

\begin{figure}[ht]
  \centerline{\includegraphics*[width=0.65\textwidth]{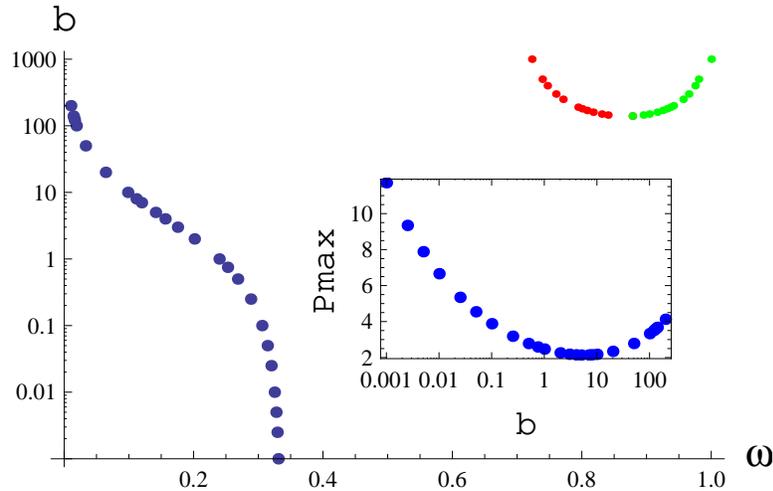}}
  \caption{Three non-communicating Brownian searchers.  The loci of
    the extrema of $P(\omega)$ in Eq. (\ref{3exp}).  Blue (green)
    circles define the position of the first (second) maximum.  Red
    circles define the position of the minimum of $P(\omega)$. The
    inset shows the maximal value $P(\omega_m)$ vs $b$ ($a$ is set equal to $1$).}
  \label{sketch7}
\end{figure}

The case $0 < \mu < 1$ is different.  Focusing on $\mu = 1/2$ (\eg,
biased BMs on a semi-infinite line), for which Eq. (\ref{k})
simplifies,
\begin{equation}
\label{3exp}
P(\omega) = \frac{4}{\pi} \sqrt{\frac{a}{b}} \frac{e^{6 \sqrt{a/ b}}}{\omega (1 - \omega) \sqrt{1 + 3 \omega}}
K_1\left(2 \sqrt{\frac{a}{b} \frac{1 + 3 \omega}{\omega (1- \omega)}}\right),
\end{equation}
we discuss a sequence of different regimes which may be observed when
$b/a$ is gradually varied, see Fig.~\ref{sketch6}.  For $b/a \ll 1$,
$P(\omega)$ is peaked at $\omega_m \approx 1/3$. For larger $b/a$,
$\omega_m$ moves towards the origin and $P(\omega_m)$ decreases (see
the inset of Fig.~\ref{sketch7}).  For yet larger $b/a$, $\omega_m$
keeps moving towards the origin but now $P(\omega_m)$ passes through a
minimum and then starts to grow.  At $b/a \approx 140$ a second
extremum emerges at $\omega \approx 0.84$ which then splits into a
minimum and a maximum (see Fig.~\ref{sketch7}) so that $P(\omega)$ becomes
bimodal.  For still larger $b/a$, the minimum moves towards $\omega =
1/2$, while the second maximum moves to $\omega = 1$.

In fact, the result in Eq.~(\ref{3exp}) can be straightforwardly
generalized for arbitrary $N$, yielding
\begin{eqnarray}
\label{generalBM}
&&P(\omega) = \frac{2 (N - 1)}{\pi} \sqrt{\frac{a}{b}} \frac{e^{2 N \sqrt{a/ b}}}{\omega (1 - \omega) \sqrt{(N - 1)^2 w + 1 - w}} \nonumber \\
&&\times K_1\left(2 \sqrt{\frac{a}{ b} \left(\frac{1}{w} + \frac{(N - 1)^2}{1 - w}\right)}\right) \;.
\end{eqnarray}
The distribution function in Eq.~(\ref{generalBM}) shows essentially
the same behavior as the one in Eq.~(\ref{3exp}); the only difference
is that the critical values of the parameter $b/a$ at which $P_{max}$
attains a minimal value or when the second maximum emerges depend on
the number of searchers $N$.

\section{Conclusions}

To conclude, in this paper we have studied the distribution
$P(\omega)$ of the random variable $\omega \sim
\tau_1/\overline{\tau}$, $\overline{\tau} = N^{-1} \sum_{k=1}^N
\tau_k$, where $\tau_k$'s are the first passage times to an immobile
target by $N$ independent searchers, which start their random motion
simultaneously from the same point in space.  Hence, $\tau_k$'s are
independent, identically distributed [with distribution $\Psi(\tau)$]
random variables.  Since $\omega$ equals, by definition, the
realization-dependent first passage time of a given searcher relative
to the realization-dependent ensemble-average first passage time of
$N$ searchers, the distribution $P(\omega)$ can be viewed as a measure
of the robustness of a given search algorithm and of the underlying
random motion (space exploration), which also probes the validity of
the \textit{mean} first passage time of a single searcher as a proper
and/or informative measure of the search efficiency.

We have considered two general forms of $\Psi(\tau)$: The one in
Eq.~(\ref{distribution}), which is appropriate for search in unbounded
domains and is characterized by a power-law long-time tail $\sim
\tau^{-1-\mu}$, where the exponent $\mu $ encodes the specific details
of the searchers' random motion, and an exponentially tempered form in
Eq.~(\ref{truncated_dist}), which is a plausible approximation for
random motion in finite domains or search assisted by a constant bias
("smell") towards the target.

We have shown that for a non-truncated distribution $\Psi(\tau)$ in
Eq.~(\ref{distribution}) with $\mu < 1$, the distribution of the
random variable $\omega$ has a characteristic $U$-shaped form so that
the most probable values of $\omega$ are $0$ and $1$.  For $N = 2$ the
distribution is symmetric around $\omega = 1/2$ with $\omega = 1/2$
being the minimum of the distribution. This signifies that the ``symmetry''
between two identical searchers is broken.  For $N > 2$ the
distribution $P(\omega)$ is skewed (by a factor $N^2$) towards the
small values of $\omega$ and can be expressed, in an explicit form, as
a one-sided $\alpha$-stable distribution with $\alpha = 2 \mu$.  For
$\mu \geq 1$, the distribution $P(\omega)$ has a bell-shaped form but
the most probable value of $\omega$ is always less than the mean value
$\langle \omega\rangle = 1/N$.  The most probable and the mean values of $\omega$
coincide only when $\mu \to \infty$.

For the exponentially truncated $\Psi(\tau)$ in
Eq.~(\ref{truncated_dist}) the distribution $P(\omega)$ always has a
bell-shaped form for $\mu \geq 1$.  For $\mu < 1$, however, the
situation is more complicated and interesting.  We realized first that
for $N = 2$ there exists some critical value of the parameter
$y_c(\mu) = b/a$, so that for $b/a < y_c(\mu)$ the distribution
$P(\omega)$ has a bell-shaped form, but for $b/a > y_c(\mu)$ it
attains an $M$-shaped form with $\omega = 1/2$ being the least
probable value and two maxima close to the edges of the
interval. This signifies that, despite the fact that $\Psi(\tau)$ has
moments of arbitrary order, two identical searchers will arrive for
the first time to the target at distinctly different
times. Consequently, in such a situation the \textit{mean} first
passage is not a proper measure of the search efficiency. We remind
that such a form of $\Psi(\tau)$ is exact for a BM taking place on a
semi-infinite one-dimensional line with a constant bias directed
towards the target.  For this physical situation, our result implies
that two identical BMs, starting at some point $x_0$ and having the same drift velocity
$v < 0$ and the same diffusion coefficient $D$, will most likely arrive
together to the target (the origin) if the Peclet number $Pe = x_0
|v|/2 D$ exceeds some critical value $Pe_c \approx 0.666$. On
contrary, if $Pe < Pe_c$, an event that these two BMs arrive
simultaneously to the target location is the \textit{least} probable
event.

Turning next to an unbiased BM on a finite interval $[0,L]$ we recall
that $b$ should be proportional to $L^2/D$ and $a \sim
x_0^2/D$. Hence, $\sqrt{a/b} \sim x_0/L$.  This suggests a somewhat
strange result that for two unbiased identical BMs on a finite 1D interval the
modality of the distribution should depend on how far is the starting
point from the reflecting boundary.
Since $\Psi(\tau)$ in Eq.~(\ref{truncated_dist}), which we have
used for the derivation of this result, is an approximate form of the
first passage time distribution, we have revisited this problem using
exact forms of $\Psi(\tau)$ for a BM in 1D, 2D and 3D
spherical domains with a reflecting boundary.  We have shown that
indeed, the very shape (modality) of the distribution $P(\omega)$
depends on the ratio $x_0/L$.

We have realized that for $x_0/L < \chi_c(d)$, where $\chi_c(d) \approx 0.61, 0.64$ and $0.68$ for 1D, 2D and 3D, respectively, $P(\omega)$ is an $M$-shaped function of $\omega$ with a minimum at $\omega = 1/2$ so that here two unbiased identical BMs
will most probably arrive to the target location for the first time at distinctly different times.
For $x_0/L > \chi_c(d)$, the distribution has a maximum at $\omega = 1/2$ so that, mathematically, the most probable event is that two BMs arrive for the first
time to the location of the target simultaneously. Note, however, that $P(\omega)$ is a "bell-shaped" function only in 1D (although is still rather broad) but in 2D and 3D,
$P(\omega)$ is nearly flat in an extended region around the maximum and rather abruptly vanishes in the vicinity of the edges of the interval. This signifies that here sample-to-sample fluctuations are very significant. In case when the starting point $x_0$ of two BMs is uniformly distributed within the domain (outside the target), we found that $P(\omega) \equiv 1$. This allows us to conclude that in neither of these well-studied situations (apart of, with some reservations, 1D case with $x_0/L \sim 1$) the mean first passage time of an individual searcher can be considered as a robust measure of the search process efficiency.

We argue that a similar behavior will take place in finite 1D systems
for fractional BM with arbitrary Hurst index $H$ or for
$\alpha$-stable L\'evy flights with $0 < \alpha < 1$, and, more
generally, for \textit{finite} systems of (not necessarily integer)
dimension $d_f$ with fractal dimension $d_w$ of random motion
trajectories given that a) $d_f \leq d_w$ (compact exploration, $\mu =
1 - d_f/d_w < 1$) or b) $d_f > d_w$ (non-compact exploration, $\mu =
d_f/d_w - 1$) but $d_f < 2 d_w$.

We have evaluated $P(\omega)$ in the case of three and more
searchers with the exponentially-truncated first passage time
distribution in Eq.~(\ref{truncated_dist}).  We have shown that for
$\mu > 1$ the distribution $P(\omega)$ is a bell-shaped function of
$\omega$ for any value of $b/a$.  Next, we have demonstrated that for
very small values of the ratio $b/a$ the distribution $P(\omega)$ has
a bell-shaped form with a maximum close to $1/N$.  Further on, we have
predicted the following sequence of regimes which can be observed upon
gradually increasing $b/a$ (the starting point $x_0$ of $N$ searchers
is moving towards the location of the target): the most probable value
$\omega_m$ of $\omega$ moves towards the origin. The value of the
maximum, $P(\omega_m)$ first decreases, passes through a minimal value
and then starts to increase. At a certain threshold value of $b/a$, a
second extremum emerges in the vicinity of $\omega = 1$, which then
splits into a minimum and a maximum so that the distribution
$P(\omega)$ becomes a skewed $M$-shaped one. This signifies, as well,
that for such a situation the \textit{mean} first passage time of a
given searcher is not a representative characteristic of the search
process.

As a final observation, we note that one may encounter a power-law distribution with a more abrupt truncation,
compared to the exponential function in Eq.~(\ref{truncated_dist}), say, a
bounded power-law (see \cite{greg2}) or a power-law tempered from both sides by a
Gaussian function:
\begin{equation}
\Psi(\tau) = \frac{\left(a \, b\right)^{\mu/2}}{K_{\mu/2}\left(2 a/b\right)} \exp\left( - \frac{a^2}{\tau^2}\right) \, \frac{1}{\tau^{1 + \mu}} \, \exp\left( - \frac{\tau^2}{b^2}\right).
\end{equation}

\begin{figure}[ht]
  \centerline{\includegraphics*[width=0.65\textwidth]{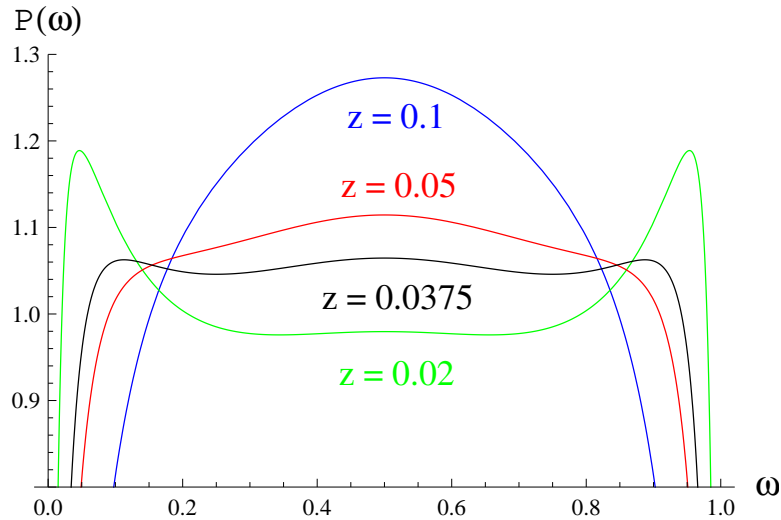}}
  \caption{$P(\omega)$ in Eq. (\ref{oufff}) for
  a faster than exponential truncation of the intermediate power-law behavior. Here the parameter $z = 2 a/b$.}
  \label{sketch9}
\end{figure}

In this case, the distribution of the random variable $\omega$ is given by
\begin{equation}
\label{oufff}
P(\omega) = \frac{1}{K_{-\mu/2}^2\left(2 a/b\right)} \, \frac{1}{\omega (1 - \omega)} \, K_{-\mu} \left(2 \frac{a \left(\omega^2 + (1 - \omega)^2\right)}{b \omega (1 - \omega)}\right).
\end{equation}
One finds that $P(\omega)$ in Eq. (\ref{oufff}) is always a bell-shaped function for $\mu \geq 1$. For $\mu < 1$, depending on the value of $z = 2 a/b$, it may have a unimodal, or a three-modal form.

\section*{Acknowledgments}

The authors wish to thank O. B\'enichou, C. Godr\`eche, S. N. Majumdar, I. M. Sokolov
and M. Vergassola for helpful discussions.



\end{document}